\documentclass[12pt]{article}
\usepackage{amsthm,amsmath,latexsym,amssymb,amsfonts,amscd}
\usepackage{graphics,lscape,fancyhdr,array,stmaryrd,euscript,wrapfig}
\usepackage{empheq,wrapfig}
\usepackage{verbatim,slashed}
\numberwithin{equation}{section}
\usepackage{hyperref,setspace}
\usepackage{tikz-cd}
\usepackage{mathrsfs}
\usepackage[numbers,sort&compress]{natbib}
\usepackage[nottoc]{tocbibind}

\newcommand{\pl}{\partial}

\newcommand{\be}{\begin{align}}
\newcommand{\ee}{\end{align}}

\newcommand{\CMax}{{C_{M}}}

\newcommand{\mm}{{\ensuremath{{\mu}}}}

\newcommand{\ga}{A}
\newcommand{\gb}{B}
\newcommand{\gc}{C}
\newcommand{\gd}{D}

\newcommand{\gad}{{A'}}
\newcommand{\gbd}{{B'}}
\newcommand{\gdd}{{C'}}

\newcommand{\bry}{{{\bar{y}}}}

\newcommand{\hs}{{\mathfrak{hs}}}

\newcommand{\fud}[2]{{}^{#1}{}_{#2}\,}
\newcommand{\fdu}[2]{{}_{#1}{}^{#2}\,}

\DeclareMathOperator{\sign}{sign}

\newcommand{\Tr}{{\mathrm{Tr}\,}}

\newcommand{\hhbar}{{\hbar}}

\newcommand{\ordB}[1]{{#1}_{(2)}}
\newcommand{\ordC}[1]{{#1}_{(3)}}
\newcommand{\aux}[1]{#1^{\text{aux}}}
\newcommand{\phys}[1]{#1^{\text{phys}}}

\newcommand{\besubeqs}{\begin{subequations}}
\newcommand{\esubeqs}{\end{subequations}}

\usepackage{tensor}
\usepackage{todonotes}

\renewcommand{\bar}[1]{\overline{#1}}

\newcommand{\sysAdS}{\ensuremath{\mathrm{HS}(\Lambda)}}
\newcommand{\sysAdSMat}{\ensuremath{\mathrm{HS}(\Lambda,\mathrm{Mat}_M)}}
\newcommand{\sysAdSMatnoone}{\ensuremath{\mathrm{HS}(\Lambda,\slashed{1},\mathrm{Mat}_M)}}
\newcommand{\sysAdSnoone}{\ensuremath{\mathrm{HS}(\Lambda,\slashed{1})}}
\newcommand{\sysFlat}{\ensuremath{\mathrm{HS}(\Lambda=0)}}
\newcommand{\sysFlatnoone}{\ensuremath{\mathrm{HS}(\Lambda=0,\slashed{1})}}
\newcommand{\sysFlatMax}{\ensuremath{\mathrm{HS}(\Lambda=0,{1}_{Max})}}
\newcommand{\sysAdSMax}{\ensuremath{\mathrm{HS}(\Lambda,{1}_{Max})}}

\usepackage[all,cmtip]{xy}

\begin{document}
\pagenumbering{gobble}
\hfill
\vskip 0.01\textheight
\begin{center}
{\Large\bfseries 
Matter-coupled higher spin gravities in $\boldsymbol{3d}$: \\ [7pt] no- and yes-go results}

\vskip 0.03\textheight
\renewcommand{\thefootnote}{\fnsymbol{footnote}}
Alexey \textsc{Sharapov},${}^{a}$
Evgeny \textsc{Skvortsov}\footnote{Research Associate of the Fund for Scientific Research -- FNRS, Belgium}\footnote{Also at Lebedev Institute of Physics}${}^{b}$ \& Arseny  \textsc{Sukhanov}${}^{b}$
\renewcommand{\thefootnote}{\arabic{footnote}}
\vskip 0.03\textheight
{\em ${}^{a}$Physics Faculty, Tomsk State University, \\Lenin ave. 36, Tomsk 634050, Russia}\\
\vspace*{5pt}
{\em ${}^{b}$ Service de Physique de l'Univers, Champs et Gravitation, \\ Universit\'e de Mons, 20 place du Parc, 7000 Mons, 
Belgium}\\ \vspace*{5pt}

\vskip 0.05\textheight

\begin{abstract}
Massless higher-spin fields show no preference for any value of the cosmological constant in $3d$. All matter-free higher-spin gravities in 3d are equivalent to Chern--Simons theories with an appropriate choice of gauge algebra. For various reasons, including holography, it is important to enrich them with matter fields. In $(A)dS_3$, the coupling of matter fields to higher-spin fields is well-known to the leading order and is determined by the representation theory. We extend this result to flat space, where the relevant higher-spin algebra is the Poisson algebra, aka $w_{1+\infty}$. However, we show that both in flat and $(A)dS_3$ spaces there are no nontrivial higher order deformations/interactions. Nevertheless, by enlarging the field content with some auxiliary fields and taking advantage of the chiral higher-spin gravity's vertices, it is possible to construct an exotic matter-coupled theory on $(A)dS_3$. It also admits a flat limit. The equations of motion have the form of a Poisson sigma-model and a meaningful action has the form of a Courant sigma-model. We also explore the potential for embedding this theory into a holographic duality.
\end{abstract}

\end{center}
\newpage
\tableofcontents
\newpage
\section{Introduction}
\setcounter{footnote}{0} 
\pagenumbering{arabic}
\setcounter{page}{1}
The paper is motivated by the great interest in $3d$ higher spin gravities over the last decade, which was mostly inspired by AdS/CFT correspondence. The progress was limited by the fact that an existing proposal \cite{Prokushkin:1998bq} to couple massless higher-spin fields to matter was found to be incomplete, see e.g. \cite{Boulanger:2015ova,Skvortsov:2015lja}, and by the puzzles on the CFT side, the so-called problem of light-states, see e.g. \cite{Gaberdiel:2010pz,Castro:2011iw,Chang:2011vka,Gaberdiel:2012uj,Perlmutter:2012ds,Campoleoni:2013lma}. As a consequence, the study was confined to either matter-free higher-spin gravities, see e.g. \cite{Campoleoni:2010zq,Henneaux:2010xg}, that can always be formulated as Chern--Simons theory \cite{Blencowe:1988gj,Bergshoeff:1989ns,Pope:1989vj,Fradkin:1989xt,Grigoriev:2019xmp,Grigoriev:2020lzu} or to the lowest order in matter fields, see e.g. \cite{Ammon:2011ua,Chang:2011mz,Lovrekovic:2018hgu}.\footnote{An exception is \cite{Kessel:2015kna}, which eventually has led to revealing in \cite{Boulanger:2015ova,Skvortsov:2015lja} that the existing proposals are incomplete. } In this paper we attempt to systematically search for matter-coupled higher-spin gravities ending up with some no-go and yes-go results.  

In $d>3$, massless higher-spin fields have propagating degrees of freedom and display slightly different features, as compared to $3d$. In $d\geq4$ (a) interactions of higher-spin fields may involve higher derivatives and all such interactions are needed for consistency \cite{Berends:1984rq,Bengtsson:1983pg, Bengtsson:1983pd,Metsaev:1991mt,Metsaev:1991nb}; (b) there are no matter-free multiplets, the smallest ones found first by Flato and Fronsdal \cite{Flato:1978qz} contains matter fields with $s=0,1/2$ in addition to the genuine massless gauge fields $s\geq1$. Initially, it was noticed in \cite{Fradkin:1986qy} that certain non-analyticity in the cosmological constant was present in the higher-spin interactions, for example, in the gravitational ones. However, as it was shown the same year \cite{Bengtsson:1986kh} there exist two-derivative gravitational interactions of higher-spin fields already in flat space. Therefore, the non-analyticity noticed in \cite{Fradkin:1986qy} is an artifact of the chosen field variables. Later, it was also shown that there is a one-to-one correspondence between cubic vertices in flat and anti-de Sitter spaces \cite{Metsaev:2018xip}. Let us also note that the known well-defined (perturbatively local) examples such as conformal higher-spin gravity \cite{Segal:2002gd,Tseytlin:2002gz,Bekaert:2010ky}, the matter-free $3d$ ones \cite{Blencowe:1988gj,Bergshoeff:1989ns,Pope:1989vj,Fradkin:1989xt,Grigoriev:2019xmp,Grigoriev:2020lzu} and Chiral theory \cite{Metsaev:1991mt,Metsaev:1991nb, Ponomarev:2016lrm,Skvortsov:2018jea,Skvortsov:2020wtf} are all smooth in the cosmological constant. 

In $3d$, massless higher-spin fields (already the graviton) do not have propagating degrees of freedom, which allows for more freedom and ambiguities. For example, (a) interactions of higher-spin fields are just one- and two-derivatives ones and they do not have all to be present for consistency, in particular, there is no big difference between lower-spin and higher-spin interactions; (b) matter fields can be decoupled and there is a great flexibility in choosing the higher-spin multiplets. Since the genuine higher-spin interactions are not that different from the usual low-spin ones there is even less argument to make with regard to the importance of the cosmological constant. There does not seem to be a single invariant statement in the literature that makes any preference for the value of the cosmological constant. 

So far the study of $3d$ higher-spins has been limited to two cases: (1) matter-free higher spin gravities, i.e., basically, to Chern--Simons theories with gauge algebras $\mathfrak{g}$ containing the gravitational subalgebra $sl_2$ in such a way that $\mathfrak{g}$ decomposes into $sl_2$ representations at least one of which is bigger than the adjoint; (2) coupling of matter to higher-spin fields to the leading order. Many results have been obtained for (1), including the study of boundary conditions, asymptotic symmetries, and higher-spin black holes, see, e.g. \cite{Campoleoni:2010zq,Henneaux:2010xg,Ammon:2012wc,Gaberdiel:2010pz,Gaberdiel:2012uj}. Much less was done for (2), see, e.g. \cite{Ammon:2011ua,Chang:2011mz,Lovrekovic:2018hgu,Kessel:2015kna}. The leading order coupling of matter fields to higher-spin fields is determined by the representation theory: matter fields must fall into a representation of the higher-spin symmetry. Therefore, this coupling does not contain any dynamical information. 

A proposal for the matter-coupled higher-spin gravity in $3d$ was made in \cite{Prokushkin:1998bq} in the form of generating equations whose solution for certain auxiliary variables should yield interactions. However, the proposal needs to be completed with a detailed recipe on how to extract the interactions. The default solution does not give interactions that are at least well defined \cite{Boulanger:2015ova,Skvortsov:2015lja}.\footnote{``Well-defined'' includes giving systematic predictions for interactions that lead to sensible physical observables, the simplest being holographic correlation functions.} An additional feature of \cite{Prokushkin:1998bq} is that the free spectrum contains, on top of the matter and higher-spin fields, some additional fields with no clear physical interpretations, which so far has not been accounted for by any $AdS/CFT$ proposal. 

In the present paper, we would like to reconsider the problem of matter-coupled higher-spin gravities in $3d$ from scratch.\footnote{Note that the proposal of \cite{Prokushkin:1998bq} has not been constructed by starting from the correct free system that would contain only higher-spin fields and matter fields. Instead, the recipe of \cite{Vasiliev:1990en} was adapted to $3d$ by dropping half of the variables with an immediate consequence that certain nonphysical fields must be present. It seems that the problem of matter-coupled higher-spin gravities in $3d$ has never been addressed systematically, but see \cite{Mkrtchyan:2017ixk,Kessel:2018ugi,Fredenhagen:2019hvb,Fredenhagen:2019lsz,Fredenhagen:2024lps} for the important first steps in this direction.} We also would like to include the flat space case, where the locality of interactions is easier to control. 

Our findings are as follows. The system with non-zero cosmological constant turns out to be easier to handle and it can be shown not to have any nontrivial deformations up to the order where the stress tensor is expected to show up. We also propose a simple free system in flat space where the relevant higher spin algebra turns out to be the Poisson algebra on $\mathbb{R}^2$ (the algebra of area-preserving diffeomorphisms, also known as $w_{1+\infty}$) and the matter fields transform in the dual module. This system can be shown to be a smooth limit of the (anti-)de Sitter one, where the higher spin algebra is the (even subalgebra of) Weyl algebra $A_1$ (quantization of the Poisson algebra) or $gl_\lambda$. This system does not have any nontrivial deformations either under certain assumptions. 

To be as general as possible, we stay at the level of equations of motion and employ the language of Free Differential Algebras \cite{Sullivan77, vanNieuwenhuizen:1982zf,DAuria:1980cmy} that was proposed in \cite{Vasiliev:1988sa} as a tool to study deformations of higher-spin systems. We study the deformations in the first two orders, which also includes the coupling via the stress-tensor. One can also replace the topological spin-one field with the Maxwell one, which does not save the day in the flat space. Therefore, given the very general and generous assumptions, we conclude that there are no matter-coupled higher-spin gravities in $3d$ neither in flat (with less certainty) nor in (anti-)de Sitter spaces. 

Despite the no-go result stated above, there is a simple amendment that allows us to find a nontrivial solution in (anti-)de Sitter space. First, we enlarge the symmetry algebra $A_1$ by tensoring it with the matrix algebra $\mathrm{Mat}_2$ as was done in \cite{Prokushkin:1998bq}. This trick may not be safe and enriches the field content with certain fields that do not have any immediate physical interpretation, but can be argued \cite{Prokushkin:1998bq} to remain topological. After this modification, we can directly take the vertices (to be precise, the $A_\infty$-algebra) of Chiral higher-spin gravity \cite{Skvortsov:2022syz, Sharapov:2022faa, Sharapov:2022awp, Sharapov:2022wpz, Sharapov:2022nps} and find a non-trivial deformation. The reason for this to work is that Chiral theory's $A_\infty$-algebra $\mathbb{A}_c$ is the tensor product $\mathbb{A}_c= \mathbb{A}\otimes B$ of a nontrivial $A_\infty$-algebra $\mathbb{A}$ that has $A_1$ as the underlying vector space and an associative algebra $B$. For Chiral theory, $B$ should be chosen to be another copy of $A_1$ or $A_1\otimes \mathrm{Mat}_N$ (if we need supersymmetry and/or Yang--Mills gaugings). Here, we choose $B=\mathrm{Mat}_2$, or, more generally, $B=\mathrm{Mat}_M$. This gives us a matter-coupled higher-spin gravity with an additional `topological' sector. 

The outline of the paper is as follows. In Section \ref{sec:over}, we briefly review matter-free higher-spin gravities and discuss how to couple matter fields to the leading order. Here we also review the old results and present some new ones about the flat space limit. In Section \ref{sec:deform}, we attempt to deform various theories and sooner or later find an obstruction. Not to end with such a discouraging result we construct a theory in Section \ref{sec:yesgo}, which is at the price of extending the field content. We end with some conclusions in Section \ref{sec:conclusions}, where we discuss possible loopholes in the arguments and attempt to embed the new theory into the holographic context by adapting the Gaberdiel--Gopakumar conjecture.

\section{Overview}
\label{sec:over}
In this section, we review the existing formulations of free higher-spin fields, interactions thereof and how to couple higher-spin fields to matter to the lowest order. Some nice and more detailed overviews can be found in \cite{Kessel:2016hld,Campoleoni:2024ced}. There are a couple of new results as well.

\subsection{Matter-free higher spin fields}
\label{sec:}
\paragraph{Very free fields.} It is customary to describe a massless  field of spin $s$ by a symmetric rank-$s$ tensor $\Phi_{a(s)}\equiv \Phi_{a_1\cdots a_s}$,\footnote{A group of symmetric indices $a_1\ldots a_s$ is abbreviated as $a(s)$. More generally, all indices in which a tensor is already symmetric or needs to be symmetrized are denoted by the same letter.} which is called Fronsdal's field
\cite{Fronsdal:1978rb} and is a natural generalization of the metric field $g_{\mu\nu}$. Therefore, it is oftentimes called the metric-like approach. The field is subject to the following free gauge symmetry: 
\begin{align}
   \delta\Phi^{a(s)}=\nabla^a \xi^{a(s-1)}\equiv \nabla^{a_1} \xi^{a_2\cdots a_s}+\text{permutations}\,.
\end{align}
Hereinafter $\nabla$ stands for the flat or (anti-)de Sitter covariant derivative.
In addition, the field is required to be double-traceless, $\Phi\fud{a(s-4)bc}{bc}=0$. 
The gauge parameter $\xi^{a(s-1)}$ is traceless, $\xi\fud{a(s-3)b}{b}=0$. Free fields satisfy the equations of motion
\begin{align}\label{adsfron}
\square \Phi^{a(s)}-\nabla^a \nabla_m\Phi^{m a(s-1)}+\tfrac12\nabla^a \nabla^a \Phi\fud{a(s-2)m}{m}-m^2\Phi^{a(s)}+2\Lambda g^{aa}\Phi\fud{a(s-2)m}{m}=0\,,
\end{align}
where $m^2=-\Lambda s(s-3)$ and $\Lambda$ is the cosmological constant. The form of the wave operator is uniquely fixed by the gauge invariance. The equations come from an action \cite{Fronsdal:1978rb,Buchbinder:2001bs}. In $3d$, the equations describe no propagating degrees of freedom. However, they may lead to complicated `boundary dynamics' in the sense of $AdS_3/CFT_2$-correspondence once interactions are introduced, the first precursor being the famous \cite{Brown:1986nw}. By `interactions' one can understand certain nonlinear gauge-invariant completions of the free actions/equations of motion.\footnote{Note that a very efficient formalism to study interactions of higher-spin fields --- the light-cone approach, see e.g. \cite{Bengtsson:1983pg, Bengtsson:1983pd,Metsaev:1991mt,Metsaev:1991nb, Metsaev:2005ar} --- cannot be applied since there are no bulk degrees of freedom. } 

It turns out that the frame-like approach
\cite{Aragone:1979hx,Aragone:1980rk,Vasiliev:1980as} is much more efficient in constructing interactions as compared to the metric-like approach reviewed here, which was first demonstrated in \cite{Blencowe:1988gj}. The Fronsdal field can be replaced by a pair of one-forms $e^{a(s-1)}\equiv e^{a_1\cdots a_{s-1}}_\mu\, dx^\mu$ and $\omega^{a(s-1)}\equiv \omega^{a_1\cdots a_{s-1}}_\mu\, dx^\mu$ that are symmetric and traceless in $a_1,\ldots, a_{s-1}$. The second-order Fronsdal equations are equivalent to a pair of the first-order equations
\begin{align}\label{masslesslorentz}
    \nabla e^{a(s-1)} +\epsilon\fud{a}{bc} h^b \wedge \omega^{a(s-2)c}&=0\,,&
    \nabla \omega^{a(s-1)} +\Lambda \epsilon\fud{a}{bc} h^b \wedge e^{a(s-2)c}&=0\,.
\end{align}
Here, $h^a$ is the background dreibein one-form, $\nabla h^a=0$. 
The Fronsdal field can be identified with the totally symmetric part of the dreibein
\begin{align}
    \Phi^{a_1\cdots a_s}&=e^{a_1\cdots a_{s-1}}_\mm h^{\mm a_s}+\text{symmetrization}\,.
\end{align}
To take even more advantage of the $3d$ situation we transfer everything into the spinorial language. Namely, thanks to the isomorphism $so(1,2)\sim sl(2,\mathbb{R})$, instead of Lorentz (spin)-tensors it is more convenient to work with tensors of $sl(2,\mathbb{R})$. A symmetric and traceless rank-$s$ tensor $T_{a_1\cdots a_s}$ becomes a rank-$2s$ tensor $T_{A_1\cdots A_{2s}}$ of $sl(2,\mathbb{R})$, a representation we denote $V_{s}$. The same equations of motion \eqref{masslesslorentz} read
\begin{align}\label{framespinor}
    \nabla e^{\ga(2s-2)}+h\fud{\ga}{\gb}\wedge \omega^{\gb \ga(2s-3)}&=0 \,,&
    \nabla \omega^{\ga(2s-2)}+\Lambda h\fud{\ga}{\gb}\wedge e^{\gb \ga(2s-3)}&=0\,.
\end{align}
The background dreibein 
$h^{\ga\gb}$ and spin-connection $\varpi^{\ga\gb}$ satisfy
\begin{align}\label{hw}
    dh^{\ga\gb}+\varpi\fud{\ga}{\gc}\wedge h^{\gb\gc}&=0\,, & d\varpi^{\ga\gb}+\varpi\fud{\ga}{\gc}\wedge \varpi^{\gb\gc}+\Lambda h\fud{\ga}{\gc}\wedge h^{\gb\gc}&=0\,,
\end{align}
which are just the flatness condition for the (anti-)de Sitter algebra ($\Lambda\neq 0$) or for the Poincare algebra  ($\Lambda=0$).   
Since the anti-de Sitter algebra is $so(2,2)\sim sl(2,\mathbb{R})\oplus sl(2,\mathbb{R})$ and the de Sitter algebra $so(1,3)\sim sl(2,\mathbb{C})$, for $\Lambda\neq0$ it is convenient to introduce\footnote{$\iota=1$ for the anti-de Sitter case and $\iota=i$ for the de Sitter one and we keep $\Lambda=\pm1$ in most of the formulas. } $A_L=\varpi+ \iota e$, $A_R=\varpi-\iota e$ and rewrite the flatness condition \eqref{hw} as
\begin{align}
    dA^{\ga\gb}_L+A_L\fud{\ga}{\gc}\wedge A^{\gb\gc}_L&=0\,, & dA^{\gad\gbd}_R+A_R\fud{\gad}{\gdd}\wedge A^{\gbd\gdd}_R&=0\,,
\end{align}
Here, we use primed and unprimed indices to distinguish between the two two-dimensional representations of the spacetimes symmetry algebra.\footnote{They are the fundamental representations of the two $sl(2,\mathbb{R})$ in the anti-de Sitter case and of the (anti)-fundamental representations of $sl(2,\mathbb{C})$ in the de Sitter one.} The free higher-spin equations are simply the covariant constancy conditions
\begin{align}\label{freemassless}
    D_L \Omega^{\ga(2s-2)}_L&=0\,, & D_R \Omega^{\gad(2s-2)}_R&=0\,,
\end{align}
where $\Omega_{L,R}=\omega \pm \iota e$, and $D_{L,R}$ are the covariant derivatives with respect to $A_L$ and $A_R$. Of course, $(D_L)^2=(D_R)^2=0$ and the gauge transformations are $\delta \Omega_{L,R}=D_{L,R}\xi_{L,R}$. 

\paragraph{Interactions.}  Note that the free higher-spin equations look like a linearization of the flatness condition for a Lie algebra, which we denote $\hs$, whose decomposition under the Lorentz algebra $sl(2,\mathbb{R})$ contains representation $V_{s-1}$. It turns out that this is always the case \cite{Grigoriev:2020lzu}: (a) all matter-free higher spin gravities in $3d$ have the form of the flatness condition for some Lie algebra $\hs$, which we call the higher-spin algebra (very far from being unique); (b) if $\hs$ admits a non-degenerate invariant form, which we loosely denote $\Tr$, then there is an action and it has the Chern--Simons form
\begin{align}
    S_{\text{CS}}[\Omega]&=\int \Tr\left[\Omega \wedge d \Omega +\frac23 \Omega\wedge \Omega\wedge \Omega\right]\,,
\end{align}
where now $\Omega$ takes values in $\hs$. As for $\hs$, one can say that everything goes, but the simplest choice is $\hs=sl(N,\mathbb{R})\oplus sl(N,\mathbb{R})$, more generally, $\hs=\mathfrak{g}\oplus \mathfrak{g}$ for some $\mathfrak{g}$. Note that the free fields as well as the background are described by a pair of connections. To reproduce the usual gravity $\hs$ has to contain the Poincare or (anti)-de Sitter subalgebra. In particular, one has to specify an embedding of the gravitational Lorentz subalgebra $sl(2,\mathbb{R})$ into $\hs$. For $\Lambda\neq0$ the action is usually taken to be the difference $S[\Omega_{L,R}]=S_{\text{CS}}[\Omega_L]-S_{\text{CS}}[\Omega_R]$ of Chern--Simons action for $\mathfrak{g}$. The latter contains the Einstein--Hilbert action, to get which taking the difference of two $sl(2)$ Chern--Simons actions is important. However, classically one can take any linear combination of two $\mathfrak{g}$ Chern--Simons actions. The spectrum of the theory is given by the decomposition of $\hs$ under the $sl(2,\mathbb{R})$ Lorentz subalgebra. For example, the principal embedding of $sl(2)$ into $sl(N)$ gives
\begin{align}
    sl(N)&=V_1\oplus V_2 \oplus \cdots \oplus V_{N-1} \,,
\end{align}
which leads to the spectrum $s=2,3,\ldots,N$. Another popular choice is $gl_\lambda$ (also called $\hs(\lambda)$), which is defined as the following quotient \cite{Feigin}:
\begin{align}
    gl_\lambda&=U(sl_2)/I_\lambda\,, && I_\lambda=U(sl_2)[\boldsymbol{C}_2+(\lambda^2-1)]\,.
\end{align}
In other words, it is the quotient of the universal enveloping algebra by the two-sided ideal generated by the `Casimir plus a constant'. 
For generic $\lambda$, it is a simple associative algebra whose decomposition under the canonical $sl(2)$ subalgebra is given by
\begin{align}
    gl_{\lambda}&= V_0\oplus V_1 \oplus \cdots
\end{align}
For $\lambda \in \mathbb{Z}$ the algebra develops a two-sided ideal $J_\lambda$ such that the quotient is $gl({|\lambda|})$, hence the name $gl_\lambda$ \cite{Feigin}. Note that $gl(N)$ decomposes as 
\begin{align}\label{masslesscase}
   gl(N)&= V_0\oplus V_1\oplus\cdots \oplus V_{N-1}\,.
\end{align}
Considered as a Lie algebra w.r.t. the commutator, $gl_\lambda$ contains the one-dimensional ideal generated by the unit, and the complement is a simple Lie algebra, called $sl_\lambda$. Thus, $gl_\lambda=u(1)\oplus sl_\lambda$. The same is true for $gl(N)$, of course.  

Another useful property of $gl_\lambda$ is that $gl_{1/2}$ is the even subalgebra $A_1^e$ of the smallest Weyl algebra $A_1$. The algebra $A_1$ can be understood as the algebra of functions $f(\hat y)$ (say, polynomials) of generators $\hat{y}_A$ that satisfy 
\begin{equation}\label{yy}
    [\hat{y}_A,\hat{y}_B]=-2\epsilon_{AB}
    \end{equation}
 in our normalization. It is the same as the algebra of functions of commuting variables $y_A$ with the product replaced by the Moyal--Weyl star-product. The $sl(2)$ generators $t_{AB}=-\tfrac14 \{\hat{y}_A, \hat{y}_B\}$ satisfy
\begin{align}
[t_{\ga\gb},t_{\gc\gd}]&=\epsilon_{\gb\gc}t_{\ga\gd}+\ldots
\end{align}
By construction, there is a natural embedding of $sl_2$ into $gl_\lambda$ and we denote the $sl_2$-generators the same way, $t_{\ga\gb}$. To summarize, the equations of motion for matter-free higher-spin gravities with $\hs=gl_\lambda \oplus gl_\lambda$ can be written as two zero-curvature conditions
\begin{align}
    dA &= A\star A\,, & dB &= B\star B 
\end{align}
where $A=A_L$, $B=A_R$ are connection of $gl_{\lambda}$ and $\star$ denotes the associative product in $gl_{\lambda}$, for which numerous forms are available \cite{Pope:1989sr, Fradkin:1990qk, Korybut:2014jza, Joung:2014qya, Basile:2016goq,Sharapov:2022phg}. For $\lambda=1/2$ we can enjoy the Weyl algebra realized as the Moyal--Weyl star-product. In the latter case, $A_{L,R}=A_{L,R}(y)=A_{L,R}(-y)$ (the projection to the even subalgebra). Note that we can optionally disregard the topological spin-one field since it takes values in the $u(1)$-ideal and does not `interact' with any other field. On the contrary, one can take the tensor product $gl_\lambda \otimes \mathrm{Mat}_M$ as a higher-spin algebra, which makes the abelian spin-one field into a nonabelian one. Still, there is the central element $1\otimes \mathbf{1}_M$ that is an abelian spin-one field that can be disregarded. 

\subsection{Matter fields}
\label{sec:}
By matter fields we mean the scalar field $s=0$ and fermion $s=\tfrac12$. For simplicity we discuss the scalar field, but everything applies mutatis mutandis to the fermion as well. The free equation is just the Klein--Gordon equation for $\phi(x)$. However, one needs to make sure that a given higher-spin algebra $\hs$ acts on $\phi$, at least when it is on-shell. This is an important selection rule. For example, none of $\hs=sl(N)$, $N>3$ is admissible. Indeed, the scalar field should transform in some unitary representation of $\hs$. The Gelfand--Kirillov dimension of $sl(N)$ is $N-1$, i.e. one needs functions of at least $N-1$ variables to have a unitary representation, while the on-shell $3d$ scalar field is a function of just $2$ variables.\footnote{An interesting option is $sl(3)$. } 

One possible solution is to take $gl_{\lambda}$ whose representation theory is completely equivalent to that of $sl(2)$. The higher-spin algebra is $\hs=gl_{\lambda} \oplus gl_{\lambda}$ as a Lie algebra. Given that for any associative algebra $A$ there is a natural bimodule structure on $A$ itself, one has two module structures on $A$ over $\mathrm{Lie}(A)$, the left and the right ones. Therefore, the simplest possible option to take a zero-form $C$ valued in $gl_\lambda$ and write
\begin{align}\label{scalarcoupled}
    d C&= A\star C-C\star B 
\end{align}
for two flat connection $A$ and $B$ of $gl_\lambda$. In case $\lambda=1/2$, we can take $C=C(y|x)=C(-y|x)$ and use the Moyal--Weyl star-product for $y^A$s. It is instructive to explain why \eqref{scalarcoupled} has anything to do with the scalar field, but see \cite{Vasiliev:1992ix,Prokushkin:1998bq,Kessel:2016hld,Campoleoni:2024ced} for more detail. First, we choose the background to be
\begin{align}
    A&= \tfrac12 (\varpi^{AB}+h^{AB}) t_{AB}\,, & B&= \tfrac12 (\varpi^{AB}-h^{AB}) t_{AB}\,,
\end{align}
where the dreibein $h^{AB}$ and spin-connection $\varpi^{AB}$ satisfy \eqref{hw} for $\Lambda=1$. The equation for the master-field $C$ reduces to
\begin{align}
    \nabla C& =\tfrac12 h^{AB} \{t_{AB}, C\}_\star \,.
\end{align}
The first two equations in the expansion over $y$ (or $t_{AB}$) read
\begin{align}\label{bloodyscalar}
    \nabla_\mu \phi& = -\tfrac12 h^{AB}_\mu C_{AB}\,, & \nabla_\mu C^{AB}&= \tfrac12 h^{AB} \phi-\tfrac12 h^{CD}_\mu C\fud{AB}{CD} \,,
\end{align}
where $C=\phi + C^{AB} t_{AB} + \tfrac12 C^{ABCD}t_{AB} t_{CD}+\ldots$. This is the Klein--Gordon equation in the first order form: $C^{AB}=2\nabla^{AB} C$ and on contracting the second equation with $h^\mu_{AB}$, one gets 
\begin{align}
    (\square -\tfrac34 \Lambda)\phi&=0\,,
\end{align}
where we temporarily reinstated $\Lambda$. Higher components of $C$ are just the auxiliary fields that are expressed by virtue of the equations of motion as symmetrized gradients of the scalar field, $C^{A(2n)}\sim \nabla^{AA}\cdots \nabla^{AA}\phi$.

\subsection{Generalizations, variations}
\label{sec:}
There is a number of simple modifications of Eq. \eqref{scalarcoupled} that are as good as, or seem even better than, Eq. \eqref{scalarcoupled} itself and should be included in the scope.

\paragraph{Simple variations.} Before all, let us refer to the system above as system-\sysAdS. Its matrix extension, when $gl_\lambda$ is replaced with $gl_\lambda\otimes \mathrm{Mat}_M$, will be referred to as  system-\sysAdSMat. Note that in the latter case, the master-field $C$ takes values in the same algebra and describes $M^2$ scalars charged under $\mathrm{Mat}_M\oplus \mathrm{Mat}_M$.\footnote{Looking at the $u(1)$-part we find that the covariant derivative acts as $D_\mu\phi=\pl_\mu \phi -A_\mu \phi+ \phi B_\mu $. Therefore, the scalar has opposite charges with respect to the two $u(1)$ gauge fields. The symmetry is reducible since the combined transformation leaves the scalar invariant.  }

Firstly, if $\lambda=1/2$ and $C$ is odd in $y$, then one finds the spin-half matter field $\psi_A(x)$ as the first component that satisfies the Dirac equation \cite{Vasiliev:1992ix,Prokushkin:1998bq}. Extension to other values of $\lambda$ requires the so-called deformed oscillators, \cite{Yang:1951pyq, Boulware1963, Mukunda:1980fv,Vasiliev:1989re}. 

Secondly, one can disregard the topological spin-one field, i.e. to take $\hs=sl_\lambda\oplus sl_\lambda$. The equations do not change, just $A$ and $B$ do not contain the unit of the algebra $gl_\lambda$ anymore. However, note that $C$ still takes values in $gl_\lambda$, which determines the module structure. Physically, the unit of $C$ as an element of $gl_\lambda$ is associated with the scalar field $\phi(x)$. Mathematically, we cannot drop the $u(1)$ part of $C$ since the Lie module structure comes from an associative one. In other words, $gl_\lambda$ is a natural Lie-bimodule over $sl_\lambda$. Let us refer to a system without the abelian spin-one fields as to system-\sysAdSnoone. Its matrix extension, again with the $u(1)$ spin-one fields dropped, can be called system-\sysAdSMatnoone. 

Finally, let us note that it may sometimes be convenient to rewrite the equations in terms of the (higher-spin) dreibein $e$ and spin-connection $\omega$ fields, which gives, see e.g. \cite{Ammon:2020fxs},
\besubeqs\label{framelike}
\begin{align}
    de&=\omega\star e +e\star \omega\,,\\
    d\omega&= \omega\star \omega +e\star e\,,\\
    dC&= [\omega,C]_\star +\{e,C\}_\star \,.\label{scalarew}
\end{align}
\esubeqs
This free system is consistent with the same options as discussed above: with or without the spin-one field and with/without the matrix extension.

\paragraph{Flat space.} As it was already discussed in the introduction, there are no facts that would prefer $\Lambda\neq0$ to $\Lambda=0$, especially in $3d$. Therefore, let us consider the flat space analog of the matter-coupled system above. We keep the same field content, but only have to modify the structure of the algebra and its action on $C$. Convenient variables for the flat limit are not the left/right connections $A$, $B$, but the (higher-spin) spin-connection $\omega=\omega(y)=\omega(-y)$ and the (higher-spin) dreibein $e=e(y)=e(-y)$, for which we use the same generating functions as for $gl_{1/2}=A_1^e$.\footnote{Yet another realization of $gl_\lambda$ is via the deformed oscillators \cite{Yang:1951pyq, Boulware1963, Mukunda:1980fv,Vasiliev:1989re}, which, after passing to symbols thereof, allows us to keep the same generating functions of $y^A$ as for $A_1$. Note, however, that the proof in Section \ref{sec:hscoh} applies to $\lambda=1/2$ only since we use the Moyal-Weyl star-product, while the general discussion in the sections before \ref{sec:hscoh} applies to generic $\lambda$. } 

There is a number of flat space systems one can define. The simplest equations have the form
\begin{subequations}\label{LInf:FreeFDA}
    \begin{align}
        d\omega&=\left\{\omega,\omega\right\}\,,\\
        d e&=\left\{\omega, e\right\}\,,\\
        dC&=U_1\left(w,C\right)+U_2\left(e,C\right)\,.
    \end{align}    
\end{subequations}
Here, $\{\bullet,\bullet\}$ is the Poisson bracket (the second term of the Moyal--Weyl star-product, the first one being the pointwise multiplication). In term of the symbols of operators it is\footnote{For any poly-differential operator $v(f_1,\ldots,f_n)$ that takes a collection of functions $f_i(y)$ to a function it is convenient to introduce auxiliary variables $y^A_i$ and set $p_i^A\equiv \pl^A_{y_i}\equiv \pl/ \pl y^A_i$. Then, any such operator can be represented by a function $v(y, p_1,\ldots ,p_n)$ whose action on $f_1(y_1)\cdots f_n(y_n)$ is followed by setting $y_i=0$. Further improvement of the notation is to define $p_0\equiv y$ and $p_i\cdot p_j \equiv -p^A_i p^B_j \epsilon_{AB}$. It may be useful to note that $\exp[y\cdot p_1]f(y_1)=f(y_1+y)$. \label{ft:polydiff} }
\besubeqs
\begin{align}
\left\{f,g\right\}=&\partial^Cf\partial_Cg=\left(p_1\cdot p_2\right)\exp[y\cdot\left(p_1+p_2\right)]f\left(y_1\right){g\left(y_2\right)|}_{y_{1,2}=0}\,.
\end{align}
\esubeqs
The first two equations imply that the higher-spin algebra $\hs$ is only a Lie algebra this time. Let $P$ be the Poisson algebra of (even) functions $f(y)$ in $y$. It acts naturally on itself through the adjoint action and $\hs= P\ltimes P$, where the second $P$ is the adjoint module. Indeed, $\{\omega,\omega\}$ implies $P$, while $\{\omega, e\}$ defines $e$ to be in the adjoint of $P$. Note that there is no `cosmological term' $\{e,e\}$ in the flat limit. $\{\omega,\bullet\}$ correctly reproduces the standard action of the spin-connection $\tfrac14\omega^{AB} y_A y_B\in \omega$.  The maps $U_1$ and $U_2$ define the action of the higher-spin algebra on $C$
\begin{align}
&U_1\left(\omega, C\right)=R\left(\omega, C\right)=p_{0}\cdot p_1\exp\left[p_{0}\cdot p_2+p_{1}\cdot p_2\right]\left.\omega\left(y_1\right)C\left(y_2\right)\right |_{y_{1,2}=0}\,,\\
&U_2\left(e, C\right)=\exp\left(p_{1}\cdot p_2+p_{0}\cdot p_2\right)e\left(y_1\right){C\left(y_2\right)|}_{y_{1,2}=0}\,,
\end{align}
where $R$ is the action of $P$ on the dual space $P^*$, which is defined via 
\begin{align}
    \left\langle\left\{f, g\right\}| C\right\rangle=\left\langle f|R\left(g,C\right)\right\rangle
\end{align}
and the pairing is
\begin{align}
    \left\langle f; g\right\rangle=\exp\left[p_1\cdot p_2\right]f\left(y_1\right){g\left(y_2\right)|}_{y_{1,2}=0}\,.
\end{align}
$U_2$ can also be explained: it is the action of the commutative algebra $C[y]$ on the dual space, effectively, $U_2(e,C)=e(\pl)C(y)$. In Appendix \ref{app:limit}, we show that the flat space system above is a limit of the $(A)dS$ one based on $gl_{1/2}$. Let us refer to the system here-above as to system-\sysFlat. We cannot turn on Yang--Mills gaugings because one cannot tensor a Lie algebra and an associative/Lie algebra, in general. Nevertheless, we can drop the spin-one fields in $\omega$ and $e$ and define the system-\sysFlatnoone. We also note that the flat system above does not come from an associative higher-spin algebra 'as usual', which, e.g., was assumed in \cite{Ammon:2020fxs}.

The simple example above can be generalized. A higher-spin analog of the Poincare algebra was defined in \cite{Ammon:2017vwt} as a limit of $gl_\lambda\oplus gl_\lambda$, see also \cite{Campoleoni:2021blr}. The net result is that the higher-spin `Lorentz' algebra is $gl_\lambda$ and the higher-spin `translations' are defined to be $gl_\lambda$ as a vector space  where the Lorentz algebra acts naturally on. More generally, we can take any associative algebra $A$ or a Lie algebra $\mathfrak{g}$ and extend it with itself as a natural bimodule ($A$-case) or module ($\mathfrak{g}$-case), i.e. we have $\hs=A\ltimes A$ and $\hs=\mathfrak{g}\ltimes \mathfrak{g}$. This construction gives sensible algebras for $gl_\lambda$, $sl_\lambda$ and $sl_N$. In order to add the matter, one can define $C$ to take values in the dual (bi)-module $\hs^*$.

\paragraph{Changing Chern--Simons to Maxwell.} One more option\footnote{We are grateful to Xavier Bekaert for a very useful discussion of this point.} is to let the spin-one field be dynamical, i.e. instead of the topological field ({\`a} la Chern--Simons) one can choose the usual Maxwell equations, which in $3d$ describes the scalar degree of freedom.\footnote{This is a natural option when extrapolating from higher dimensions to $d=3$ since in $4d$ and higher the massless spin-one does have propagating degrees of freedom and is a part of the standard higher-spin multiplet. } That the field equations are equivalent to the Klein--Gordon equation can be made manifest
\begin{align}
    dF&=0 && d^*F=0 && \Longleftrightarrow && F=*d\phi && d*d\phi=\square \phi=0 \,,
\end{align}
where $F=dA$ and $d^*=*d*$ with $*$ being the Hodge dual. $F=*d\phi$  solves the Maxwell equations $d^*F=0$, but the Bianchi identities $dF=0$ turn into the Klein--Gordon equation with vanishing mass (irrespective of the curvature). We can cast this idea into the FDA form as follows. Since the degrees of freedom are those of a scalar, we need to add a zero-form $\CMax$ with exactly the same equations as for the scalar field. However, we have to set $\lambda=1$ to describe a massless scalar in $(A)dS_3$. In order to relate the gauge potential to the first derivative of the scalar, we can write
\begin{align}\label{maxwell}
    dA&= e_{AC} \wedge e\fdu{B}{C}\, C^{AB}\,.
\end{align}
Recall that $C_{AB}\sim h_{AB}^\mu \pl_\mu \phi$ on-shell, \eqref{bloodyscalar}. The last equation has to be a projection of a certain $\mathcal{V}(e,e,C)$-cocycle that respects the higher-spin symmetries. We will discuss two systems \sysAdSMax{} and \sysFlatMax{} with a self-evident encoding.

The system with a scalar field and a Maxwell one results from the direct extrapolation of the higher-spin algebra and its associated field spectrum (which is given by the Flato--Fronsdal theorem) from $d=4$ to $d=3$. Interestingly, the simplest higher-spin algebra in $AdS_3$ is based on $gl_{1/2}$, while the extrapolation of the results from $d=4$ dimension requires $\lambda=1$. Note that $\lambda=1$ leads to the $W$-symmetry on the boundary that can be realized by the free boson CFT, as it is the case in higher dimensions. Note that $A_1^e\sim gl_{1/2}$ admits the usual oscillator 'singleton' realization and, for that reason, also appears special. Nevertheless, there does not seem to be anything special about $\lambda=1/2$ algebraically and we consider this point generic.

In a bit more detail, let us take the construction of Eastwood \cite{Eastwood:2002su} that defines the higher-spin algebra as the symmetry algebra of the free conformal scalar field in $d$-dimensions or, equivalently, as the quotient of $U(so(d,2))$ by a certain two-sided ideal. Let us denote the algebra $\mathcal{E}_d$. We use the fact that $so(2,2)\sim sl(2)\oplus sl(2)$ and, hence, $U(so(2,2))\sim U(sl_2)\otimes U(sl_2)$. Denoting the generators of the two $sl(2)$-algebras $t_{1,2}^{AB}$, a generic element of the universal enveloping algebra is 
\begin{align}
    f&= \sum_{n,m} f_{A(2n)|B(2m)}\, t_1^{AA}\cdots t_1^{AA} \otimes t_2^{BB}\cdots t_2^{BB}\,.
\end{align}
The ideal defined in \cite{Eastwood:2002su} is generated by the degree-two elements, see e.g. \cite{Campoleoni:2021blr},
\begin{align}
    t_1^{AA} \otimes t_2^{AA}&\sim0\,, &t_1^{AA}t_{1\,AA}^{\vphantom{1}} &\sim0\,, &t_2^{AA}t_{2\,AA}^{\vphantom{1}} &\sim0\,, \\
    t_1\fud{A}{B} \otimes t_2^{AB}&\sim0\,,&t_1^{AA}\otimes t_{2\,AA}^{\vphantom{1}} &\sim0\,.
\end{align}
The algebra $\mathcal{E}_2$ looks almost like $gl_1\oplus gl_1$,\footnote{Indeed, the $sl(2)$ Casimir vanishes for $\lambda=1$ and the generator of the two-sided ideal in $U(sl_2)$ is $t_{AA}t^{AA}\sim0$. All the other relations above imply $t_1^{AB}\otimes t_2^{CD} \sim0$. } but it is not. The crucial difference from $gl_1\oplus gl_1$ is that there is just one unit, while in $gl_1\oplus gl_1$ we would have $(1,0)$, $(0,1)$ and the actual unit $(1,1)$. There is no canonical way to identify the units of two associative algebras $A$, $B$ in $A\oplus B$. What happens is something special: due to the fact that $t_{i}^{AA} t_{i\, AA}\sim 0 $, $i=1,2$, none of the units of the two $U(sl_2)$ subalgebras of $U(sl_2)\otimes U(sl_2)$ can appear in the product of two generators. In other words, if we denote the structure constants of $gl_1$ by $f_{IJ}^K$, $I,J,K=0,1,2,...$ with $I=0$ corresponding to the unit of the algebra, then $f_{IJ}^{K=0}=0$. For generic $\lambda$ algebra $gl_\lambda$ is simple as an associative algebra and decomposes into $u(1)\oplus sl_\lambda$ as a Lie algebra. In case $\lambda=1$, one observes that $gl_1=k\oplus sl_1$ as an associative algebra ($k=\mathbb{R}$ or $\mathbb{C}$). Therefore, the value $\lambda=1$ is very special. 

Let us take $C$ with values in $\mathcal{E}_1$. It can be decomposed as 
\begin{align}
    C&= \phi + \sum_{k\geq 1} C_{A(2k)}^1 t_1^{AA}\cdots t_1^{AA} + \sum_{k\geq 1}C_{A(2k)}^2 t_2^{AA}\cdots t_2^{AA}= \phi+C_1(t_1)+C_2(t_2)\,.
\end{align}
We write the usual equation \cite{Vasiliev:1992gr}
\begin{align}\label{tad}
   dC&= \Omega \star C-C\star \tilde{\Omega}\,, 
\end{align}
where $\Omega=\tfrac12 (\varpi+h)t_1 +\tfrac12 (\varpi-h)t_2$ is the background connection and $\tilde f(t_1,t_2)=f(t_2,t_1)$ is an automorphism of the algebra that flips the sign of the translations $P\sim t_1-t_2$ and leaves the Lorentz generators $L\sim t_1+t_2$ invariant. It gives \eqref{scalarew} upon $\varpi\rightarrow\omega$ and $h\rightarrow e$, the latter is related to \eqref{scalarcoupled}. The fact that the unit can never be generated in the product of two nontrivial elements implies that $\phi$ decouples from everything, $d\phi=0$. For the rest of the fields we find
\begin{align}
       dC_{1,2}&= [\varpi,C_{1,2}]_\star \pm\{h,C_{1,2}\}_\star \pm 2 h \phi\,.
\end{align}
where $\pm$ corresponds to $C_{1,2}$ and the equations are essentially identical. Even though the scalar $\phi$ decouples, the rest of the zero-forms could describe two propagating scalar degrees of freedom. Each system begins with
\begin{align}
    t_{AA}&: & \nabla C^{AA}&= \zeta_2  h_{BB}C^{AABB}+ h^{AA} \phi\,,
\end{align}
where $\zeta_2\neq0$. Here, $C^{AA}$ is the dual $C^\lambda=\epsilon^{\mu\nu\lambda} F_{\mu\nu}$ of the Maxwell field strength $F_{\mu\nu}$ written in the spinorial language, $C_{AA}=\sigma_{AA}^\lambda C_\lambda$. For $\zeta_2\neq0$ the equations imply the Maxwell equations $\nabla_\mu F^{\mu\nu}=0$ and the Bianchi identities get modified into $dF=*\phi$, i.e. $\nabla_\mu C^\mu=\phi$ or $\pl_{[\mu} F_{\nu\lambda]}=\epsilon_{\mu\nu\lambda}\phi$, not to forget that $\pl_\mu \phi=0$. Since $\phi=\text{const}$ it does not contribute to the propagating degrees of freedom. For $\lambda=1$, we can also consider $C$ taking values in the two-sided ideal to be quotiented out to get $sl_{\lambda=1}$. In other words, we can drop the unit from $C$. More generally, we observe
that\footnote{This is very easy to see for $\lambda=1/2$ where $y_A\star f= (y-\pl_A)f$ in the Weyl ordering and, hence, $y_{A}y_{A}\star f= (y_Ay_A-2y_A\pl_A +\pl_A\pl_A)f$, which agrees with the three terms in \eqref{usualt}.}
\begin{align}\label{usualt}
    t_{AA} \overbrace{t_{BB}...t_{BB}}^{n} \sim t_{(AA} t_{BB}...t_{BB)} + \bullet \epsilon_{A(B}t_{AB}t_{BB}...t_{BB)} +\zeta_n \epsilon_{AB}\epsilon_{AB} t_{BB}...t_{BB}\,.
\end{align}
Importantly, $\zeta_{n>1}\neq 0$ and $\bullet$ is some irrelevant coefficient that does not vanish as well. It is only $t_{AA}t^{AA}=0$, hence, $\zeta_1=0$. Given that the Maxwell field requires $\lambda=1$ and there is no simple realization of  $gl_{\lambda=1}$ we refrain from considering this option.

\subsection{Interactions?}
\label{sec:interactions}

\paragraph{Problem.} Eq. \eqref{scalarcoupled} and the like couples the scalar field to the higher-spin background. In particular, it couples it to gravity in the canonical way (minimal coupling). However, this cannot be the end of the story because there must be a stress-tensor on the r.h.s. of Einstein's equations built out of the scalar field and one also expects some higher rank stress tensors to show up on the r.h.s. of `Fronsdal's equations' that are hidden inside the FDA. Therefore, the flatness conditions need to be deformed to have a physically sounding theory. This is the problem we address in the paper.  Note that without including the backreaction the theory is non-Lagrangian since the minimal coupling to gravity that is present in the $C$-equation has to induce the stress-tensor on the r.h.s. of $d\omega=\ldots$ equation, but the latter is absent.\footnote{We discuss the property of being (non-)Lagrangian for a theory whose equations of motion are equivalent to the ones we have. It is clear that the covariant constancy equation $dC=...$ is non-Lagrangian by itself (more equations than fields), even though it is equivalent to the Klein-Gordon equation, which admits an action. }

\paragraph{Some attempts to introduce interactions.} Let us review what else is known about interactions of higher-spin fields with matter. At least to the lowest order one can couple the matter field $\phi$ to higher-spin fields via the usual Noether coupling, which gives schematically the action
\begin{align}\label{Noether}
S= S_{CS} +S_{KG}+ \sum _s a_s\int J(\phi,\phi)_{\ga(2s)} \Phi^{\ga(2s)}=S_{CS}+S_{KG}+\int \Tr \ast J(\phi,\phi) \wedge e\,.
\end{align}
Here, $J_{A(2s)}=\phi \nabla^s \phi+\ldots $ are the conserved higher-spin currents constructed from $\phi$ and $S_{CS}$, $S_{KG}$ are the Chern--Simons and the Klein--Gordon actions, respectively. The currents can be coupled to the higher-spin fields, which can be done either with the Fronsdal fields or with the frame-like field $e$ that contains them. The coupling is invariant to the lowest order as well as the identification of the Fronsdal fields as the totally symmetric components of $e$ works only at the free level, see e.g.
\cite{Campoleoni:2012hp,Fredenhagen:2014oua}. The relative strength $a_s$ of couplings can be read off from the higher-spin algebra, see \cite{Kessel:2015kna,Lovrekovic:2018hgu}. Let us note that the Noether coupling does make higher-spin currents (including the stress-tensor) contribute to the r.h.s. of the free Fronsdal equations (including the free Einstein's equation), but it is not obvious how to extend it to include arbitrary higher-spin backgrounds. Therefore, equations \eqref{framelike} and action \eqref{Noether} are somewhat complementary to each other.  

It should also be mentioned that one can study various interaction vertices of massless higher-spin fields with themselves and with the matter fields \cite{Mkrtchyan:2017ixk,Kessel:2018ugi,Fredenhagen:2019hvb,Fredenhagen:2019lsz,Fredenhagen:2024lps}. Here, the results have been encouraging in the sense that there are many nontrivial vertices that can be used to construct a theory.

\section{Deformation problem}
\label{sec:deform}
In this section, we discuss less optimistic results. Let us first summarize the initial data. In the case of $(A)dS_3$ we have two flat connections $A$, $B$ and a scalar field $C$ in the bi-fundamental of $gl_\lambda$ or $sl_\lambda$ or in the matrix extensions thereof. As discussed $C$ takes values in $gl_\lambda$ in both cases. The difference between $gl_\lambda$ and $sl_\lambda$ is that the former features an abelian topological spin-one field. The fields $A$, $B$, $C$ are subject to the field equations, given by the ABC-system \sysAdS{} or \sysAdSMax{} of the form
\begin{align}\label{initial}
    dA&= A\star A\,, & dB&= B\star B\,, & 
    d C&= A\star C-C\star B \,.
\end{align}
When $A$ and $B$ occupy just the $sl(2)$ subalgebra of $\hs$, a nondegenerate solution describes the anti-de Sitter geometry and the last equation imposes the Klein--Gordon equation. The most general ansatz for cubic vertices reads
\begin{equation}\label{ABC}
    \begin{array}{rcl}
        dA &= &A\star A +V_0(A,A,C)+V_1(A,C,B)+V_2(C, B,B)+\cdots \,,\\[3mm]
        dB & =&B\star B+W_0(C,B,B)+W_1(A,C,B)+W_2(A,A, C)+  \cdots\,,\\[3mm]
        dC &=&A\star C-C\star B+U_1(A,C, C)-U_2(C,C,B)+\cdots\,.
    \end{array}
\end{equation}
Here, dots on the right stand for vertices of order four and higher. If the fields are decorated with matrix factors we need to take different orderings of the fields into account, e.g. ${V}(A,A,C)$ becomes a sum of three different structures ${V}_1(A,A,C)$, ${V}_2(A,C,A)$ and ${V}_3(C,A,A)$.\footnote{The idea to write down higher-spin equations in the form of $d\Phi=Q(\Phi)$, where $\Phi$ is some set of forms and $Q$ does not involve any spacetime derivatives explicitly dates back to \cite{Vasiliev:1988sa}. The mathematical structure behind ($Q$) is known as Free Differential Algebras (FDA) \cite{Sullivan77, vanNieuwenhuizen:1982zf,DAuria:1980cmy} and is a particular case of $L_\infty$-algebras, see also the beginning of Section \ref{sec:hscoh}. } 

We will consider two cases below: there is an abelian spin-one field and without the abelian spin-one field. In the former case it is quite easy to show that no nontrivial deformation exists and the proof does not rely on the details of the higher-spin algebra apart from the fact that there is a unit and, hence, it covers $\Lambda\neq0$ and $\Lambda=0$. In the latter case, we could not avoid a detailed analysis of the cohomology of the higher-spin algebra and restrict ourselves to the generic point $\lambda=1/2$. 

\subsection{Abelian spin-one}
\label{sec:}

Once an abelian (Chern--Simons) spin-one field is in, it is not hard to see that the free equations (\ref{initial}) enjoy no non-trivial interaction in the form of an FDA. The proof below relies on the  fact that the algebra $\hs$ is unital and no other details will matter, i.e. the proof applies to $gl_\lambda$, $gl_\lambda\otimes \mathrm{Mat}_N$ and to the flat space system with the spin-one field under which the matter is charged.

Component-wise, $A=A_0dx^0+A_1dx^1+A_2dx^2$ and $B=B_0dx^0+B_1dx^1+B_2dx^2$ are one-forms with values in $\hs$. 
Notice that the interaction vertices in (\ref{ABC}), being two forms, are completely determined by their values on the fields of the form $A=A_1dx^1+A_2dx^2$ and $B=B_1dx^1+B_2dx^2$.\footnote{Indeed, the structure maps/vertices $\mathcal{V}$, ... do not involve spacetime derivatives and act only on the (fiber) indices hidden in the $y$-dependence. Therefore, in order to recover the entire map, say $\mathcal{V}(A,A,C)$, it suffice to know it on $\mathcal{V}(A_1,A_2,C)\, dx^1\wedge dx^2$.  } Therefore, we are free to impose some conditions on the zero components $A_0$ and $B_0$ of the form fields. 

Formal integrability of the partial differential equations (\ref{ABC}) implies a sequence of quadratic relations to be imposed on the interactions vertices. All of them originate from the nilpotency of the exterior differential, $d^2=0$. Relations that depend on different sets of fields are obviously independent of each other. The only relation of the $A^3C$ type reads 
$$ 
\Omega _0: = V_0(A,A,C)\star A-A\star V_0(A,A,C)+V_0(A\star A,A,C)-V_0(A,A\star A,C)+V_0(A,A,A\star C)=0.
$$
 Let us set $A_0=1$ or, what is the same, $i_{\partial_0}A=1$. Then
\begin{equation}
    i_{\partial_0}\Omega_0=V_0(A,A,C)+A\star R(A,C)-R(A\star A,C)+R(A,A\star C)+R(A,C)\star A=0\,,
\end{equation}
where $R(A,C)=2V_0(1,A,C)$ and we used the fact that $V_0(a,b,C)=-V_0(b,a,C)$ for any functions $a$ and $b$ with values in $\hs$. This allows us to express every integrable cubic vertex $V_0$ in terms of some bilinear map $R(A,C)$. Clearly, all such vertices are trivial and can be removed by the field redefinition $A\rightarrow A+R(A,C)$.  
If we now set $V_0=0$, then the integrability condition for the vertex $V_1$ takes the form
\begin{equation}
    \Omega_1:=V_1(A,C,B)\star A-A\star V_1(A,C,B)+V_1(A\star A,C,B)-V_1(A,A\star C,B)=0\,.
\end{equation}
Again, taking $A_0=1$ and $B_0=0$, we find
\begin{equation}
    i_{\partial_0}\Omega_1=V_1(A,C,B)+A\star S(C,B)-S(A\star C, B)+S(C,B)\star A=0\,,
\end{equation}
where $S(C,B)=V_1(1,C,B)$. 
The last equation says the the vertex $V_1$ is trivial and can be removed by the  field redefinition $A\rightarrow A+S(C,B)$. 
To complete the analysis of the first equation in (\ref{ABC}), we set $V_0=V_1=0$ and write the integrability condition for the remaining cubic vertex $V_2$:
\begin{equation}
    \Omega_2:=V_2(C,B,B)\star A -A\star V_2(C,B,B)+V_2(A\star C, B,B)=0\,.
\end{equation}
As an immediate consequence, we get
\begin{equation}
    i_{\partial_0}\Omega_2=V_2(C,B,B)=0\,.
\end{equation}

Now, one can repeat the above arguments word for word to conclude that all $W$ vertices are trivial as well. 
It remains to consider $U$ vertices. Under assumption that all $V$ and $W$ are zero, we obtain the two integrability conditions
of $A^2C^2$ and $B^2C^2$ types:
\begin{equation}
\begin{array}{c}
    \Omega_3:= -A\star U_1(A,C,C)-U_1(A,C,A\star C)-U_1(A,A\star C, C)+U_1(A\star A, C,C)=0\,,\\[3mm]
     \Omega_4:= U_2(C,C, B)\star B +U_2( C\star B,C,B)+U_2( C, C\star  B, B)-U_2(C,C, B\star B)=0\,.   \end{array}
\end{equation}
Setting $A_0=B_0=1$, we find
\begin{equation}
\begin{array}{c}
    i_{\partial_0}\Omega_3= U_1(A,C,C)+A\star P(C,C)-P(A\star C, C)-P(C,A\star C)=0\,,\\[3mm]
     i_{\partial_0}\Omega_4= U_2(C,C, B) +Q(C,C)\star B -Q(C\star B, C)-Q(C,C\star B)=0\,,   \end{array}
\end{equation}
where $P(C,C)=U_1(1, C,C)$ and $Q(C,C)=U_2(C,C,1)$. 
This allows us to remove the cubic $U$ vertices by the shift $C\rightarrow C+P(C,C)+Q(C,C)$. 

Repeating the above analysis order by order shows that all interaction vertices in (\ref{ABC}) are trivial. This conclusion could have been reached in a much more economical way by applying the Hochschild--Serre spectral sequence \cite[Ch 1, \S 5]{Fuks}. 
Indeed, since the higher-spin algebra $\hs$ is unital, the associated commutator Lie algebra $[\hs]$ splits into the direct sum $[\hs]=\hs'\oplus\mathfrak{c}$, where the one-dimensional center $\mathfrak{c}$ is generated by $1\in \hs$. If $M$  is a $[\hs]$-module, then the central ideal $\mathfrak{c}\subset [\hs]$ gives rise to a Hochschild--Serre spectral sequence $E_2^{p,q}\Rightarrow H^{p+q}([\hs],M)$ with the second page $E_2^{p,q}=H^q(\hs', H^p(\mathfrak{c}, M))$. The spectral sequence converges to the Chevalley--Eilenberg cohomology of the Lie algebra $[\hs]$ with coefficients in $M$. If we further assume that $1\in\mathfrak{c}$ acts by the identity transformation of $M$, then all groups $H^p(\mathfrak{c}, M)$ vanish  and so do the cohomology groups $H^m([\hs], M)$.  It remains to note that the equations $\Omega_0=\cdots=\Omega_4=0$ considered above have the form of a cocycle condition for the cohomology of the Lie algebra $[\hs]$ with an appropriate module of coefficients. The absence of nontrivial cocycles implies the absence of nontrivial interaction vertices. 

A heuristic argument is that since the scalar $\phi\in C$ is charged under the spin-one fields of $A$ and $B$, it is impossible to construct the deformations of the r.h.s. of the form $AAC$ that is gauge-invariant. Therefore, an interesting system can be the one with two scalars $C$ and $\bar{C}$
\begin{align}\label{initialbar}
    dA&= A\star A\,, & dB&= B\star B\,, &  &
    \begin{aligned} 
        d C&= A\star C-C\star B \,, \\
        d \bar{C}&= B\star \bar{C}-\bar{C}\star A \,,
    \end{aligned}
\end{align}
It does not have cubic deformations for the same reasons, but one can imagine some $\mathcal{V}(A,A,C,\bar{C})$-type vertices to appear at the quartic order. Charged scalar fields also appeared in the initial version of $AdS_3/CFT_2$ higher-spin duality \cite{Gaberdiel:2010pz}.

\paragraph{Maxwell spin-one.} The above arguments are also applicable to the cases where the topological spin-one field is replaced by the Maxwell field under which the scalar is assumed to be charged. Indeed, it requires the same zero-form $C$ and the Maxwell equations require a nontrivial $\mathcal{V}(A,A,C)$-cocycle to get Eq. \eqref{maxwell}. Note that for the argument above one can replace $gl_{1/2}\sim A_1^e$ by $gl_\lambda$ at a generic point $\lambda$ (recall that the Maxwell field requires $\lambda=1$).

\subsection{Higher spin cohomology}
\label{sec:hscoh}
Let us come back to the system \eqref{ABC} and exclude the spin-one field, which makes the proof from the previous section inapplicable. It is a particular case of a sigma-model, where the source $Q$-manifold $\mathcal{M}$ is the parity shifted tangent bundle $\Pi T M$ of our spacetime manifold $M$ and the target $Q$-manifold $\mathcal{N}$ is some nonnegatively graded $Q$-manifold. The algebra of functions on $\mathcal{M}$ is the (graded) commutative algebra of differential forms on $M$ and the $Q$-structure is the exterior derivative $d$. Denoting the coordinates on $\mathcal{N}$ by $\Phi^\alpha$, the associated fields $\Phi^\alpha(x,dx)$ are degree-preserving maps from $\mathcal{M}$ to $\mathcal{N}$. The condition that the two $Q$-structures are compatible with each other gives the system of PDEs
\begin{align}
    d\Phi^\alpha&=Q^\alpha(\Phi)=\sum_k Q^\alpha_{\beta_1\cdots \beta_k} \Phi^{\beta_1}\cdots\Phi^{\beta_k}\,.
\end{align}
All field theories and, more generally, PDEs can be cast into such a form, as a matter of principle, see e.g. \cite{Barnich:2009jy}. We restrict ourselves to systems with $Q^\alpha=Q^\alpha_\beta=0$. Such $Q$-structures are usually called minimal. 
Then, $Q^\alpha_{\beta_1\beta_2}$ are structure constants of some (graded) Lie algebra $\mathfrak{g}$. It is easy to see that $Q^\alpha_{\beta_1\beta_2\beta_3}$ is a $3$-cocycle of the Chevalley--Eilenberg cohomology of the algebra $\mathfrak{g}$ with coefficients in $\mathfrak{g}$. This cocycle should be nontrivial to give a nontrivial cubic vertex.  If $Q^\alpha_{\beta_1\beta_2\beta_3}=0$, then it is $Q^\alpha_{\beta_1\beta_2\beta_3\beta_4}$ that defines a $4$-cocycle and a quartic interaction vertex,  and so on. Therefore, the deformation problem, i.e. the consistency of various vertices, amounts to computing the Chevalley--Eilenberg cohomology of a given higher-spin algebra.

For the ABC system \eqref{ABC}, the target manifold $\mathcal{N}$ is supported in degrees $0$ and $1$, the corresponding coordinates and fields denoted by $C$ and $A$, $B$, respectively. Here, $\mathfrak{g}$ is the Lie algebra $\hs=sl_\lambda \oplus sl_\lambda$ extended trivially with its module $C$. Nontrivial cubic vertices are in one-to-one correspondence with elements of the cohomology group $H^3(\mathfrak{g},\mathfrak{g})$. If this last group is zero, then we should study the next cohomology group $H^4(\mathfrak{g},\mathfrak{g})$ to find out whether there are any nontrivial quartic vertices. 

Since $\hs=sl_\lambda \oplus sl_\lambda$, we can first compute the cohomology of $sl_\lambda$. This is equivalent to setting $B=0$ or, equivalently, one can set $A=0$. More specifically, we set $\lambda=1/2$ as to take advantage of the Moyal--Weyl star-product realization. Depending on the type of the vertex we consider, cocycles with values in different modules are needed. The zero-form $C$ forms the left module, i.e. $Q(C)=A\star C$. The vertices $U$ take values in the left module as well. The $V$-vertices take values in the adjoint module, i.e. $Q(V)=A\star V- (-)^{|V|} V\star A$. The $B$-vertices $W$ take values in the trivial module (with respect to $A$), i.e. $Q(W)=0$. Lastly, $Q(A)=A\star A$. Putting all this together, we find, for example, for the $U$ vertices the following action:
\begin{align}
\begin{aligned}
        Q &U(A,\ldots,A,C,\ldots,C)=\sum \pm U(A,\ldots,A\star A,A,\ldots,A,C,\ldots,C)+\\
    &\pm\sum U(A,\ldots,A,C,\ldots,A\star C,C,\ldots,C)-A\star U(A,\ldots,A,C,\ldots,C) \,.
\end{aligned}
\end{align}
Accordingly, we have three types of complexes $\mathcal{C}_{\text{left}}$, $\mathcal{C}_{\text{ad}}$, $\mathcal{C}_{\text{tr}}$ for the left, adjoint and trivial actions, respectively. The corresponding cohomology groups are denoted by $H^\bullet_{\text{left}}$, $H^\bullet_{\text{ad}}$, $H^\bullet_{\text{tr}}$. In practice, we are interested in cohomology in degrees $0$, $1$, $2$ and with one or two $C$ arguments. We use superscripts $q$ and $n$ on $H^{q,n}$ to indicate the degree and the number of $C$'s, respectively. We find the following results for the cohomology groups:\footnote{There does not seem to be any general methods that would allow one to compute the Chevalley--Eilenberg cohomology of infinite-dimensional algebras and even for the commutator Lie algebra of the Weyl algebra the result does not seem to be known. Nevertheless, it is easy to automatize the computation of cohomology by representing cochains as $sl_2$-invariant poly-differential operators and, then, Taylor-expanding the cochains, which reduces the problem to a finite-dimensional one, the only limitation being that we have to bound the number of components of $A$, $C$ and of the output argument. Strictly speaking, this does not allow us to calculate $H$ exactly.  On the other hand, if there is no deformation for spins $3$, $4$, \ldots, $s$ it is hard to imagine there is one for $s+1$. }
\besubeqs
\begin{align}\label{1}
    H_{\text{left}}^{q,n}&= \begin{cases}
        \mathbb{R}\,,& q=0\,, n=1\\
        0\,, & (q,n)=(0,2)\,, (1,1)\,, (1,2)\,,
    \end{cases}\\
    H_{\text{ad}}^{q,n}&=0\,, \qquad q=0,1,2\,, \quad n=1,2\,,\\
    H_{\text{tr}}^{q,n}&=0\,, \qquad q=0,1,2\,, \quad n=1,2\,.
\end{align}
\esubeqs
The only nonzero group in (\ref{1}) is generated by $U(C)=C$. 
Therefore, we observe that there are no nontrivial deformations of the equations for $A$ and $B$. The only possibility to take advantage of the nontrivial cohomology is to have some $U(C,B)$ that represent an element of $H^{0,1}_{\text{left}}$ with respect to $A$, but then it has to belong to $H^{1,1}_{\text{right}}$ with respect to $B$, the latter group being empty (most importantly, we do not even have vertices of type $U(C,B)$ since they belong to the free equations, which are already fixed). 

Since $\lambda=1/2$ is a generic point, we do not expect any deformation for other generic values of $\lambda$. Adding the matrix extension should not change anything in the analysis.\footnote{Assuming the system has a matrix extension for matrices of any size, the problem can effectively be reduced to the Hochschild cohomology. We compare the $3d$ case to the what happens in $d>3$ in Appendix \ref{app:Hoch}.} 

\paragraph{Flat space?} It would be interesting to have a closer look at the flat space case, which is based on the Poisson algebra, since the Poisson algebra is known to have a rather tricky cohomology \cite{gel1972cohomology}.  On one hand, it is hard to imagine that if there is no deformation in $(A)dS_3$ it can still be something nontrivial in flat space (also, all $AdS/CFT$ applications are out of reach in this case). Nevertheless, this is an interesting possibility to explore in the future. In addition, one of the options for a higher-spin algebra in flat space is just the Poisson algebra and it is less constraining as compare to the Weyl algebra.

\section{Chiral higher spin gravity in three dimensions}
\label{sec:yesgo}
The title is a bit of an oximoron since Chiral theory lives in $4d$, but it is possible to use the algebraic structure behind it to construct something in $3d$. The core element of the Chiral theory is a certain $A_\infty$-algebra $\mathbb{A}$.\footnote{We will not need a precise definition, more details can be found e.g. in \cite{Sharapov:2022nps}. For all that matters now, $A_\infty$-algebra is a collection of multilinear maps $m(\bullet,\ldots,\bullet)$ from a graded vector space $V$ into itself. It is important that $m$'s do not have any symmetry over the arguments. When a $Q$-structure, see the beginning of Section \ref{sec:hscoh}, is expanded near a stationary point one finds an $L_\infty$-algebra. The structure maps of an $L_\infty$-algebra are graded symmetric. Graded symmetrization of the structure maps of an $A_\infty$-algebra gives an $L_\infty$-algebra, which is exactly the procedure to get the vertices/equations we use.} Let $A_1$ be the smallest Weyl algebra, i.e. an associative algebra of polynomials in $y_A$ that satisfy the commutation relation $[y_A,y_B]_\star=-2\hbar\,\epsilon_{AB}$. $A_\infty$-algebra $\mathbb{A}$ is built on a graded vector space $V=V_{0}\oplus V_{-1}$ that is supported at degrees $0$ and $1$. Here, $V_{-1}\sim A_1$ and $V_0 \sim A_1^*$ as vector spaces.\footnote{Whenever a cyclic $A_\infty$-algebra is built on an associative algebra $A$ and its dual bimodule $A^*$, it is called a pre-Calabi--Yau algebra, see e.g. \cite{Kontsevich:2021fhr}. $\mathbb{A}$ turned out to be a pre-Calabi--Yau algebra \cite{Sharapov:2022wpz}. In more detail, since there is a canonical pairing between $A$ and $A^*$, scalars $\langle a_0, m(a_1,\ldots, a_n) \rangle$ can be formed with the help of the canonical pairing $\langle\bullet,\bullet\rangle$ between $A$ and $A^*$ and then require the structure maps to be related via cyclic permutations of the arguments in the scalars. One can think of Pre-Calabi--Yau algebras as noncommutative analogs of  Poisson structures.} The main idea is that $\mathbb{A}$ is good enough to describe on-shell fields in $3d$ given the functional dimension of the arguments it is defined on, which are functions $f(y)$ in two variables $y_1$ and $y_2$. An additional simple modification is to tensor $\mathbb{A}$ with a matrix algebra. Let us explain the construction in more detail.

\subsection{Higher spin algebra and free equations}
\label{sec:}
The field content will be bigger than in the previous sections, which is a price to pay for having a nontrivial deformation. The (associative) higher-spin algebra is $\hs=gl_{1/2}\otimes \mathrm{Mat}_2$ (the even subalgebra of the Weyl algebra tensored with the matrix algebra). Following \cite{Vasiliev:1992ix,Prokushkin:1998bq}, it is convenient to represent $\mathrm{Mat}_2$ as the Clifford algebra $\mathrm{Cl}_{2,0}$ generated by 
\begin{align}
    \{\phi,\psi\}&=0\,, && \phi^2=1\,, && \psi^2=1\,.
\end{align}
The fields are a one-form $\omega$ and a zero-form $C$. They satisfy the usual free equations
\begin{align}\label{usualones}
    d\omega&= \omega\star \omega\,, & dC&=\omega\star C-C\star \omega\,.
\end{align}
The fields can be decomposed as
\besubeqs
\begin{align}
C&= \aux{C}(y,\phi)+\psi\, \phys{C}(y,\phi)\,,\\
\omega&= \phys{\omega}(y,\phi)+\psi\, \aux{\omega}(y,\phi)\,.
\end{align}
\esubeqs
The Clifford element $\phi$ allows us to write two orthogonal projectors
\begin{align}
    \Pi_{\pm}&= \frac12 ( 1\pm \phi)\,, && \Pi_{\pm}\Pi_{\pm}=\Pi_{\pm}\,, && \Pi_\pm \Pi_\mp=0\,,
\end{align}
which is in accordance with $so(2,2)\sim so(2,1)\oplus so(2,1)\sim sl(2,\mathbb{R})\oplus sl(2,\mathbb{R})$. The generators of the anti-de Sitter algebra are realized as $P_{AB}=\rho \phi\, t_{AB}$ and $L_{AB}=t_{AB}$, where $t_{AB}=-\tfrac1{2\hbar} y_A y_B$ are the generators of $sl(2)$. Here, $\rho$ is related to the cosmological constant. The generators satisfy\besubeqs
\begin{align}
[L_{\ga\gb},L_{\gc\gd}]&=\epsilon_{\gb\gc}L_{\ga\gd}+\epsilon_{\ga\gc}L_{\gb\gd}+\epsilon_{\gb\gd}L_{\ga\gc}+\epsilon_{\ga\gd}L_{\gb\gc}\,,\\
[L_{\ga\gb},P_{\gc\gd}]&=\epsilon_{\gb\gc}P_{\ga\gd}+\ldots\,,\\
[P_{\ga\gb},P_{\gc\gd}]&=\rho^2\epsilon_{\gb\gc}L_{\ga\gd}+\ldots \,.
\end{align}
\esubeqs
Now, the trick is that $\{\phi,\psi\}=0$ implies that $[P_{\ga\gb}, \psi f]=-\psi\{P_{\ga\gb}, f \}$, i.e. the elements proportional to $\psi$ transform in the twisted-adjoint representation \cite{Barabanshchikov:1996mc,Prokushkin:1998bq}, while those independent of $\psi$ transform in the adjoint representation of the `small' higher-spin algebra $gl_{1/2}$.\footnote{The same can be done for $gl_{\lambda}$.} Let us take as a background $\Omega= \tfrac12 P_{AB} h^{AB}+\tfrac12 L_{AB}\varpi^{AB}$, so that  $d\Omega=\Omega\star \Omega$. The linearized equations \eqref{usualones} decompose into the two groups 
\besubeqs
\begin{align}
    D \phys{\omega}&=0\,, & 
    \widetilde{D}\phys{C}&=0\,,\\
    \widetilde{D}\aux{\omega}&=0\,, &
    D \aux{C}&=0\,,     
\end{align}
\esubeqs
where we introduced two types of covariant derivatives 
\besubeqs
\begin{align}\label{covTw}
\widetilde{D} f&
\equiv \nabla f+\tfrac12h^{\ga\ga} \{P_{\ga\ga}, f\}\,,\\
D g&\equiv \nabla g-\tfrac12h^{\ga\ga} [P_{\ga\ga}, g ]\,.\label{covAd}
\end{align}
\esubeqs
The equations in the first line describe free massless higher-spin fields and matter. The interpretation of the fields in the second line is less clear. In principle, we were looking for a higher-spin theory whose linearized field content is given by the first line only and failed to find one. Equation $D \aux C=0$ can be seen to describe Killing tensors, including the constant. Interpretation of $\widetilde{D}\aux{\omega}=0$ is unclear, see also \cite{Kessel:2015kna}. If we take all the fields seriously, we should find a strange theory where higher-spin fields and matter interact with Killing tensors and $\aux{\omega}$. Also, note that the field content is doubled due to $\phi$, e.g. there are two scalar fields.\footnote{A reality condition to truncate the spectrum down to just one scalar was proposed in \cite{Arias:2016ajh}. } Nevertheless, this seems to be the only option at present. The extra fields were dubbed auxiliary in \cite{Prokushkin:1998bq} and we stick to this terminology instead of calling them redundant or unphysical. Our aim is to construct a local higher-spin gravity with this field content.

\subsection{Vertices}
\label{sec:}
The content of this section is a short summary of \cite{Sharapov:2022awp,Sharapov:2022wpz}, see also \cite{Sharapov:2022nps,Sharapov:2023erv}. We will not give the exact rules to generate all the vertices, but just summarize the main points. The equations we are looking for have the following form:
\besubeqs\label{eq:chiraltheory}
\begin{align} 
    d\omega&= \mathcal{V}(\omega, \omega) +\mathcal{V}(\omega,\omega,C)+\mathcal{V}(\omega,\omega,C,C)+\ldots\,,\\
    dC&= \mathcal{U}(\omega,C)+ \mathcal{U}(\omega,C,C)+\ldots \,.
\end{align}
\esubeqs
The multilinear maps (vertices) on the r.h.s. can be shown to satisfy the $L_\infty$-relations, which are also required by the formal integrability and gauge invariance of the equations. The $L_\infty$-structure maps are obtained via graded symmetrization of certain $A_\infty$-maps. For example, 
\besubeqs\label{symmm}
\begin{align}
    \mathcal{V}(\omega,\omega,C)=\mathcal{V}_1(\omega,\omega,C)+\mathcal{V}_2(\omega,C,\omega)+\mathcal{V}_3(C,\omega,\omega)    \,, \\
    \mathcal{U}(\omega,C,C)=\mathcal{U}_1(\omega,C,C)+\mathcal{U}_2(C,\omega,C)+\mathcal{U}_3(C,C,\omega)   \,.
\end{align}
\esubeqs
Each of the $A_\infty$-maps, say $\mathcal{V}(f_1,\ldots ,f_n)$ with the $f$'s replaced eventually either by $\omega$ or $C$, has the factorized form 
\begin{align}
    \mathcal{V}(f_1,\ldots ,f_n)&= \mathcal{V}(f_1'(y),\ldots , f_n'(y)) \otimes f_1''\ast\cdots  \ast f_n''  \,,
\end{align}
where $f_i=f_i'(y) \otimes f_i''$, $f''_i\in B$, and $B$ is any associative algebra with product denoted $\ast$. For Chiral theory, $B=A_1[\bry]$ and for the $3d$ theory we take $B=\mathrm{Mat}_2$. The $\ast$-product will be implied below, but not written out explicitly, that is, $f_i(y)$ are understood to be two-by-two matrices that are multiplied in the usual way. The first few structure maps are
\begin{align}\label{hsalgebra}
    \mathcal{V}(f,g)&= \exp{[p_{01}+p_{02}+\hhbar\, p_{12}]}f({y}_1)\, g({y}_2)\Big|_{{y}_i=0} \equiv (f\star g)(y)\,,
\end{align}
which is just the Moyal--Weyl star-product, and\footnote{The notation for poly-differential operators was introduced in footnote \ref{ft:polydiff}.}
\begin{align}
    \mathcal{U}(\omega,C)&=\mathcal{U}_1(\omega,C)+\mathcal{U}_2(C,\omega)
\end{align}
with
\begin{equation}
    \begin{split}
        &\mathcal{U}_1(\omega,C)=+\exp{[\hhbar\, p_{01}+ p_{02}+p_{12}]}\, \omega({y}_1)\, C({y}_2)\Big|_{\bar{y}_i=0}\,,\\
        &\mathcal{U}_2(C,\omega)=-\exp{[p_{01}-\hhbar\, p_{02}-p_{12}]}\, C({y}_1)\, \omega({y}_2)\Big|_{{y}_i=0}\,.
    \end{split}
\end{equation}
Here, $\hbar$ is a free parameter that is present in the canonical commutation relations $[y_A,y_B]_\star=-2\hbar\,\epsilon_{AB}$. In order to have the $sl_2$-subalgebra of the higher-spin algebra realized by $t_{AB}=-\tfrac1{2\hbar}y_Ay_B$ we need $\hbar\neq0$. Just to give an all-order example, let us write the simplest map with both $\omega$'s at the leftmost positions:
\begin{align}
    \mathcal{V}_1(\omega,\omega, C,\ldots, C)= p_{ab}^n \int \exp\Big[ (1-\sum_i u_i) p_{0a} +(1-\sum_i v_i) p_{0b} +\sum_i u_i p_{a,i}+\sum_i v_i p_{b,i}+ \notag\\
      +\hhbar\, \Big(1+\sum_i (u_i-v_i) +\sum_{i,j} u_iv_j \sign(j-i) \Big ) p_{ab} \Big]\omega(y_a)\omega(y_b) C(y_1)\cdots C(y_n)\Big|_{y_{a,b,i}=0}\,.\label{allorder}
\end{align}
Integration is carried out over some compact domain in $\mathbb{R}^{2n}$ parameterized by the $u$'s and $v$'s. Other structure maps have a similar form.

\subsection{Perturbation theory}
\label{sec:}
Let us make some general observation on the structure of interaction. It is easy to see that at order $n$ in the zero-form $C$ the vertices have the following schematic form:
\begin{align}
    \mathcal{V}(\omega,\omega,C,\ldots ,C)&= (p_{ab})^n \exp\Big[\ast p_{0a}+\ast p_{0b}+ \ast \hhbar \, p_{ab} +\sum_{1\leq i\leq n} \ast p_{ai}+\sum_{1\leq i\leq n} \ast p_{bi}\Big]\,,
\end{align}
which is a sketchy form of \eqref{allorder} with the integral sign dropped, and
\begin{align*}
   \mathcal{U}(\omega,C,\ldots, C)&=\mathcal{U}(p_0,p_a,p_1, \ldots, p_{n+1})\, \omega(y_a) C(y_1)\cdots C(y_{n+1})\Big|_{y_{a,i}=0}=\\
   &\qquad \quad (p_{0a})^n \exp[\ast \hhbar\, p_{0a}+\sum_{1\leq i\leq n+1} \ast p_{0,i}+\sum_{1\leq i\leq n+1} \ast p_{a,i}]\,.
\end{align*}
Here, $\ast$ stands for some polynomial functions of the coordinates on the configuration space. The structure of the vertices implies locality. Indeed, $C^{A(2k)}\sim \nabla^{AA}\cdots \nabla^{AA}\phi$ from the free equations for the scalar field and the presence of any $p_{ij}$ in the exponent that contracts two zero-forms would imply nonlocality. Such terms are luckily absent.

It is instructive to see what the vertices above evaluate to on the purely gravitational background, which is the starting point of AdS/CFT calculations. Let $\varpi=\tfrac 12 t_{AB}\varpi^{AB}$, $h=\tfrac 12 \phi\, t_{AB} h^{AB}$ carry the spin-connection and the dreibein; together they form the background connection $\Omega=\varpi+h$. Also, let $\omega$, $C$ be generic functions, e.g. $C=C(y,\phi,\psi)$, and we use $\omega_0 \psi$, $C_0 \psi$ or $\omega_0$, $C_0$ when $\omega_0=\omega_0(y,\phi)$, $C_0=C_0(y,\phi)$ to stress that the field is linear in $\psi$ or independent of $\psi$. We find
\besubeqs
\begin{align}
    \mathcal{V}(\Omega, \Omega)&= \hbar  p_{01} p_{02} p_{12}\, \Omega(y_1) \wedge \Omega(y_2)\,, \label{coCosmo}\\
    \mathcal{V}(\varpi, A)+\mathcal{V}(A, \varpi)&= 2 \hbar  p_{01}  p_{12}e^{p_{02}}\, \varpi(y_1) \wedge A(y_2)\,,\label{coAd}\\
    \mathcal{V}(h, \omega_0\psi)+\mathcal{V}(\omega_0\psi, h)&=  \left(p_{01}^2+\hbar ^2 p_{12}^2\right)e^{p_{02}}\, h(y_1) \wedge \omega_0(y_2)\psi\,,\label{cotAd}\\
    \mathcal{U}_1(\varpi, C)+\mathcal{U}_2(C, \varpi)&=2 \hbar  p_{01}  p_{12}e^{p_{02}}\, \varpi(y_1) C(y_2)\,,\label{coAdC}\\
    \mathcal{U}_1(h, C_0\psi)+\mathcal{U}_2(C_0\psi, h)&=  \left(p_{01}^2+\hbar ^2 p_{12}^2\right)e^{p_{02}}\, h(y_1) C_0(y_2)\psi\label{cotAdC}
\end{align}
\esubeqs
for the bilinear vertices. The interpretation of these expressions is as follows. Eq. \eqref{coCosmo} gives the nonabelian terms $\omega\fud{A}{C}\wedge \omega^{BC}$, $h\fud{A}{C}\wedge h^{BC}$. Eq. \eqref{coAd} is the standard action of the spin-connection. Eq. \eqref{cotAd} gives the twisted covariant derivative, as in \eqref{covTw}. If $A$ is $\psi$-independent we just get \eqref{coAd} as in \eqref{covAd}. Eqs. \eqref{coAdC} and \eqref{cotAdC} are identical to \eqref{coAd} and \eqref{cotAd}, but it may not be obvious since they come from different vertices. It is clear that we have to take $\hbar \neq0$. 

For the trilinear vertices we find
\besubeqs
\begin{align}
    \mathcal{V}(\varpi, \varpi, C)&\sim \hbar p_{12}^2\, \varpi(y_1) \wedge \varpi(y_2)C(y_3) =0\,,\label{triLorA}\\
    \mathcal{V}(h, \varpi, C)&=0\,, \label{triLorB}\\
    \mathcal{V}(\varpi, A, C)&=0\,, \label{triLorBB}\\
    \mathcal{V}(h, h, C_0)&\sim  \hbar p_{12}^2\,h(y_1) \wedge h(y_2)C_0(y_3)=0 \label{triNoWeyl}\,,\\[2mm]
    \mathcal{V}(h, h, C_0\psi)&= \tfrac{1}{2} p_{12} \left(p_{01}+p_{13}\right) \left(p_{02}+p_{23}\right)\, h(y_1) \wedge h(y_2)C_0(y_3)\psi\,, \label{triSource} 
    \end{align}and
    \begin{align}
    \mathcal{V}(h, \omega_0, C_0)&=0\,,\label{triSourceA}\\
    \mathcal{V}(h, \omega_0\psi, C_0)\,, ~\mathcal{V}(h, \omega_0, C_0\psi)\,,~\mathcal{V}(h, \omega_0\psi, C_0\psi)&\neq0\,,\label{triSourceB}\\
    \mathcal{U}(\varpi, C, C)&=0\,, \label{triLorC}\\
    \mathcal{U}(h, C_0, C_0 )&=0\,,\label{trihCC}\\
    \mathcal{U}(h, C_0 \psi, C_0 )\,,~\mathcal{U}(h, C_0 \psi, C_0 \psi)\,,~\mathcal{U}(h, C_0, C_0 \psi)&\neq0\,.\label{triSourceC}
\end{align}
\esubeqs
Note, that $\mathcal{V}(\bullet,\bullet,\bullet)$ is the sum \eqref{symmm} of three different structures, the same applies to $\mathcal{U}(\bullet,\bullet,\bullet)$. When evaluating the vertices, we need to replace $\omega$ with $\varpi+h+\ldots$ and pick out appropriate terms. For example, we find the expressions
\begin{align}
    \mathcal{V}_1(\varpi, \varpi, C)+\mathcal{V}_2(\varpi, C, \varpi)+\mathcal{V}_3(C,\varpi, \varpi)\,,\\
    \mathcal{V}_1(\varpi, h, C)+\mathcal{V}_2(\varpi, C, h)+\mathcal{V}_3(C,\varpi, h)+\mathcal{V}_1(h, \varpi, C)+\mathcal{V}_2(h, C, \varpi)+\mathcal{V}_3(C,h, \varpi)\,.
\end{align}
Finally, the fields need to be brought to the same order and in doing so certain sign factors may contribute due to $\psi h=-h \psi$. 

Eqs. \eqref{triLorA}, \eqref{triLorB}, \eqref{triLorBB}, \eqref{triLorC} imply that the spin-connection does not appear outside the covariant derivative. Eq. \eqref{triNoWeyl} means that there are no sources to free higher-spin fields coming from the matter fields, as it should. However, Eq. \eqref{triSource} is a linear source for the unphysical one-forms from the matter fields. It could have been a problem, but for the same reason as in \cite{Vasiliev:1992ix} it can be redefined away.\footnote{It would be interesting to adjust the vertices via some field-redefinition as to eliminate this source and to see what is the effect on higher vertices. } Eq. \eqref{triLorC} implies \eqref{trihCC}. Eqs. \eqref{trihCC} and \eqref{triSourceA} imply that there is no source bilinear in the physical fields. On the contrary, Eqs. \eqref{triSourceB}, \eqref{triSourceC} produce a source whenever at least one field with $\psi$ is involved.

It is clear that $\mathcal{V}_n$ and $\mathcal{U}_n$ vanish identically on the purely gravitational background $\Omega=\varpi+h$ for $n\geq2$ due to $p_{ab}^n$ and $p_{0a}^n$, respectively. This has a number of simple consequences. Firstly, the spin-connection appears only as a part of the covariant derivative $\nabla = d+[\varpi,\bullet]$, i.e. the theory has manifest local Lorentz invariance: 
\besubeqs
\begin{align}
    \mathcal{V}(\varpi,\varpi,\overbrace{C,\ldots ,C}^n)&=0\,, && n\geq1\,, \\
    \mathcal{V}(\varpi,\omega,\overbrace{C,\ldots ,C}^n)&=0\,, && n\geq1\,, \\
    \mathcal{U}(\varpi,\overbrace{C,\ldots,C}^n)&=0\,, && n\geq2\,.
\end{align}
\esubeqs
Secondly, the purely gravitational background does not contribute to the equations, since
\begin{align}
    \mathcal{V}(\Omega,\Omega,\overbrace{C,\ldots,C}^n)&=0\,, && n\geq2\,.
\end{align}
Let us recall that the fields decompose as
\begin{align}
C&= \aux{C}(y,\phi)+\psi\, \phys{C}(y,\phi)\,, &
\omega&= \phys{\omega}(y,\phi)+\psi\, \aux{\omega}(y,\phi)\,.
\end{align}
Already at the free level, there is some source for the auxiliary fields from the matter fields \eqref{triSource}, i.e. $\aux{\omega}\neq0$. We can either choose to eliminate it via some field-redefinition or to keep it as is. In any case $\aux \omega=\aux\omega(\phys C)$ and, on top of that, $\widetilde{D}$ has a nontrivial cohomology in this sector, see \cite{Kessel:2015kna}. $\aux C$ can simply be set to zero. At the second order, we find
\besubeqs
\begin{align} 
    D\ordB{\omega}&= \mathcal{V}(\omega, \omega) +\mathcal{V}(h,\omega,C)\,,\\
    D\ordB{C}&= \mathcal{U}(\omega,C)+ \mathcal{U}(h,\omega,C)\,,
\end{align}
\esubeqs
where the second order fluctuations are marked with $\ordB{\bullet}$, but the first order fluctuations $\omega$ and $C$ are kept simple. This gives the following pattern for the mixture of physical and auxiliary fields:
\besubeqs
\begin{align} 
    D\phys{\ordB{\omega}}&= \mathcal{V}(\phys{\omega}, \phys{\omega})+ \mathcal{V}(\aux{\omega}, \aux{\omega}) +\mathcal{V}(h,\aux{\omega},\phys{C})+\mathcal{V}(h,\phys{\omega},\aux{C})\,,\\
    D\aux{\ordB{\omega}}&= \mathcal{V}(\phys{\omega}, \aux{\omega}) +\mathcal{V}(h,\phys{\omega},
    \phys{C})+\mathcal{V}(h,\aux{\omega},\aux{C})\,,\\
    D\phys{\ordB{C}}&= \mathcal{U}(\phys{\omega},\phys{C})+\mathcal{U}(\aux{\omega},\aux{C})+\mathcal{U}(h,\aux C,\phys C)\,,\\
    D\aux{\ordB{C}}&= \mathcal{U}(\phys{\omega},\aux{C})+\mathcal{U}(\aux{\omega},\phys{C})+\mathcal{U}(h,\aux C,\aux C)+\mathcal{U}(h,\phys C,\phys C)\,.
\end{align}
\esubeqs
We are happy to set $\aux C=0$ at the leading order to find
\besubeqs
\begin{align} 
    D\phys{\ordB{\omega}}&= \mathcal{V}(\phys{\omega}, \phys{\omega})+ \mathcal{V}(\aux{\omega}, \aux{\omega}) +\mathcal{V}(h,\aux{\omega},\phys{C}) \,,\\
    D\aux{\ordB{\omega}}&=\mathcal{V}(\phys{\omega}, \aux{\omega}) +\mathcal{V}(h,\phys{\omega},
    \phys{C})\,,\\
    D\phys{\ordB{C}}&= \mathcal{U}(\phys{\omega},\phys{C})\\
    D\aux{\ordB{C}}&= \mathcal{U}(\aux{\omega},\phys{C})+\mathcal{U}(h,\phys C,\phys C)
\end{align}
\esubeqs
However, $\aux C$ gets activated at the second order. Another somewhat unexpected feature is the vanishing of the stress-tensor contribution since $\mathcal{V}(h,h,C_0\psi,C_0\psi)=0$. However, $\mathcal{V}(\phys \omega,\phys \omega ,\phys C, \phys C)\neq0$ for a generic field configuration and represents the backreaction from the matter sector to higher-spin fields, which requires $\phys \omega$ with genuine higher-spin components. Compared with the Noether procedure discussed in Section \ref{sec:interactions}, we see that the theory looks non-Lagrangian. Indeed, $\mathcal{U}(\phys \omega, \phys C)$ follows from the standard current interaction; see, e.g. \cite{Kessel:2015kna}. The same interaction produces the stress-tensor contribution to the Einstein equations, but there is no such term in the $D\phys\omega$-equation. The subtlety here is that $\aux \omega\neq0$ and, being expressed in terms of $\phys C$ and $D\ordB{\phys\omega}$, contains $\mathcal{V}(h,\aux\omega, \phys C)$ which is bilinear in $\phys C$ and could effectively reproduce the stress tensor. 

At the third order, one term that prevents us from setting the auxiliary fields to zero is $\mathcal{V}(\phys{\omega},\phys{\omega},\phys{C}) \sim \psi$, which sources $\aux{\ordC{\omega}}$ and does not vanish. Another obstruction is the vertex $\mathcal{U}(\phys{\omega},\phys{C},\phys{C})$. The opposite is also true: the auxiliary fields can produce sources for the physical fields. A curious fact is that it is possible to build a collection of Killing tensors from massless higher-spin fields and matter fields, e.g. $\mathcal{U}(\phys{\omega},\phys{C},\phys{C})$ is a cocycle with respect to the free equations of motion. This does not seem to be in contradiction with unitarity because the adjoint representation of the higher-spin algebra is non-unitary.

Therefore, the auxiliary fields should be taken seriously. This is to be expected since the $A_\infty$-algebra $\mathbb{A}$ is nontrivial and $\mathrm{Mat}_2$ is simple, i.e. there cannot be any way to extract a perturbatively local theory for just physical fields in accordance with the no-go in Section \ref{sec:deform}. However, one can choose to solve for the auxiliary fields as to get some effective theory for the physical ones.

Since the auxiliary sector appears to be unavoidable, it might be interesting to discuss further generalizations of this theory. One can introduce supersymmetry and/or Yang--Mills gaugings, which is done via tensoring with another factor $\mathrm{Mat}_M$. Regardless of the latter options, one can just extend $\mathrm{Mat}_2$ to $\mathrm{Mat}_N$ with $N>2$ and, with the help of the Clifford algebra interpretation, identify $P_{AB}= \gamma_1 t_{AB}$. This cannot introduce any other types of fields at the free level. Depending on the indices of  $\gamma_{i_1\cdots i_k}$ one finds either physical or auxiliary fields.

\paragraph{Flat limit.} The cosmological constant originates from the relations $[t,t]=t$ and $\phi^2=1$, where $t\equiv t_{AB}$. These give $[\phi t, \phi t]=t$.  Therefore, in order to get the Poincare algebra it suffices to impose $\phi^2=0$ and we still need $\{\phi, \psi\}=0$. A question is what to require for $\psi^2$. Option $\psi^2=0$ gives a nilpotent algebra and would radically reduce the interactions since $(\phys{C})^2=0$. Therefore, it seems more interesting to keep $\psi^2=1$. One way to realize such an algebra is to take the Grassmann algebra in one variable $\theta$, $\theta^2=0$ and to tensor it with the Clifford algebra $\mathrm{Cl}_2$ generated by $\gamma_{1,2}$. Then $\phi=\theta \gamma_1$ and $\psi=\gamma_2$. Since one can realize $\theta^2=0$ with the help of the matrix algebras, whether we have $\phi^2=1$ or $\phi^2=0$ may be a matter of embedding $\phi$ into $\mathrm{Mat}_N$. For any $N$ one can interpret the theory as a higher-spin gravity in $3d$. 

\paragraph{Poisson/Courant sigma-model.} The $A_\infty$-algebra $\mathbb{A}$ behind Chiral theory is special, it is a pre-Calabi--Yau algebra of degree $2$. The latter can be thought of as a non-commutative analog of a Poisson structure. The point is that the symmetrization of a pre-Calabi--Yau algebra gives an $L_\infty$-algebra that defines (or is defined by) a Poisson bivector. The same also applies to its tensor product $\mathbb{A}\otimes \mathrm{Mat}_N$. Since $\omega$ and $C$ take values in $A$ and its dual $A^*$, respectively, with $A=gl_{1/2}$ we can denote the components as $\omega_i$, $C^j$. Then, the equations of motion have the form of those of the Poisson sigma-model
\begin{align}\label{PSM}
    d\omega_k&=\tfrac12\partial_k\pi^{ij}(C)\,\omega_i\,\omega_j\,, &   dC^i&=\pi^{ij}(C)\,\omega_j\,.
\end{align}
The vertices can be summed over $C$ to reveal a Poisson bivector $\pi^{ij}(C)$ on the space coordinatized by $C^i$. The data of free theory are associated with the leading term in the expansion $\pi^{ij}(C)=\pi^{ij}_k C^k +\ldots$. If we were in $2d$, the Poisson sigma-model's action \cite{Ikeda:1993fh,Schaller:1994es} would also be an action for the higher-spin gravity, but we are not. One possibility is just to introduce Lagrange multipliers for the equations, i.e., a one-form $\omega^k$ and a two-form $B_i$ to write
\begin{align}
    S[B,\omega,C]&= \int \omega^k\big(d\omega_k-\tfrac12\partial_k\pi^{ij}(C)\,\omega_i\,\omega_j\big)+ B_i \big(dC^i-\pi^{ij}(C)\,\omega_j\big)\,.
\end{align}
This is an example of a Courant sigma-model, see e.g. \cite{Chatzistavrakidis:2024utp}. 
Even though we are not aware of any scenario where just multiplying all equations by the Lagrange multipliers solves all problems, it is easy to see that the spectrum of the extended theory is sane enough. Indeed, dual one-forms $\omega^k$ satisfy the same $D\omega^k=0$ free equations as fields $\omega_k$. The two-forms do not introduce new propagating degrees of freedom as well.\footnote{There are other works on the action principle. The papers \cite{Prokushkin:1999gc,Bonezzi:2015igv} are within the setup of \cite{Prokushkin:1998bq}. The model of \cite{Fujisawa:2013lua} is similar to the proposal above, but lacks any higher vertices in the equations of motion; its main part is $\int B\wedge DC$, where $B$ is a two-form. }

\section{Conclusions and discussion}
\label{sec:conclusions}

One of the main results of the paper is that the simplest possibility to have a matter-coupled higher spin gravity in $3d$ is not realized. As we discussed, one cannot couple $sl(N)$, $N>3$, Chern--Simons theory to matter fields. We should take the $gl_{1/2}$ ($sl_{1/2}$) or, more generally, $gl_\lambda$ ($sl_\lambda$) multiplet of higher-spin fields. The scalar and fermion fields form a representation of this algebra, which can be encoded in the covariant constancy equation for a zero-form $C$. While mathematically consistent, this system is physically incomplete since the scalar field does interact with gravity (and higher spins) but there is no backreaction onto the gauge sector. It describes matter fields freely propagating on a higher-spin background. 
However, it turns out that these initial data do not allow nontrivial deformations in either of the two cases:
\begin{itemize}
    \item  a topological spin-one field is in, which can be realized with the algebra $gl_{\lambda}$ or in flat space;  
    \item   the gauge algebra is $sl_{1/2}\oplus sl_{1/2}$, which is the case of $(A)dS_3$.
\end{itemize}   
Since $sl_{1/2}$ is a generic point of $sl_\lambda$, the same should apply to $sl_\lambda$ save for the critical values of $\lambda$ and $\lambda=1$ where the Maxwell field can be added under which the matter must not be charged. 

One possible way out is to enlarge the field content till it allows for consistent interactions. One of the first consequences of the Noether procedure applied to higher-spin theories is that one cannot get away with a finite set of higher-spin fields in $d\geq4$, see e.g. \cite{Berends:1984rq}. The right multiplet of the gauge algebra is an important input. Matter fields take values in $gl_\lambda$ as a vector space. The $3d$ higher-spin algebra $gl_\lambda$ ($sl_{\lambda}$) has two types of representations: the left and the right (which translates into adjoint and twisted-adjoint), both of which can be realized on one- and zero-forms. They lead to physical and auxiliary\footnote{Let us stress that \textit{auxiliary fields} just refers to another representation of the higher-spin algebra and does not mean that the fields are expressed trough the physical ones. } fields both in one- and zero-form sectors. Once an appropriate set of fields is taken that necessarily contains both representations for one-forms as well as for zero-forms, it is possible to introduce interactions. Mathematically, one needs to take the $A_\infty$-algebra $\mathbb{A}$ of Chiral higher-spin gravity and tensor it with the matrix algebra $\mathrm{Mat}_M$, the smallest nontrivial option being $M=2$. 

The equations of motion have the form of a Poisson sigma-model due to $\mathbb{A}$ being a pre-Calabi--Yau algebra. This leads to an interesting option to write down an action by introducing the fields dual to the equations of motion, the resulting action having the form of the Courant sigma-model. While the action contains even more fields they do not seem to be worse than the already introduced auxiliary fields. 

It is hard at present to identify the precise obstruction that determines the extension of the minimal multiplet and calls for the auxiliary fields since the problem is somewhat ill-posed: to look for an a priori unknown extension of the multiplet among all representations of the higher-spin algebra such that higher order vertices exist. At the heuristic level, one sees that the $3d$-case is about $\mathfrak{g}\oplus \mathfrak{g}$ and $d>3$ is about $\mathfrak{g}$, where $\mathfrak{g}$ is some simple associative (or just Lie) algebra, while the zero-forms transform in accordance with $A\star C-C\star B$ or $A\star C-C\star A$, respectively. In the $3d$ case, different vertices 'talk' less to each other (since $A$ and $B$ are different) as compared to $d>3$ one and, as a result, there is no nontrivial solution.

The results raise an important question for AdS/CFT correspondence: what is the interpretation of the auxiliary fields on the CFT side? From the field theory point of view the content looks bizzare: in addition to the fields we are interested in we also find a multiplet of Killing tensors and a multiplet of one-forms with yet unknown field-theory interpretation. Nevertheless, this seems to be the only possibility for now. Therefore, it would be important to give a CFT interpretation of these additional fields. 

The following simple and naive modification of the initial $AdS_3/CFT_2$ proposal \cite{Gaberdiel:2010pz,Gaberdiel:2012uj} seems to resolve some of the issues, not without introducing new ones. The appearance of the unphysical fields is caused by an attempt to apply the usual `gravity standards'. The pure gravity is an $sl_2\oplus sl_2$ gauge theory and the standard boundary conditions \cite{Brown:1986nw} lead to two (anti-)holomorphic copies of the Virasoro algebra.\footnote{This does not seem to be related to various chiral CFT/gravity proposals, see e.g. \cite{Grumiller:2013at} for a review.} Coming to higher spins, we cannot select a gravitational $sl_2\oplus sl_2$-subalgebra and define a theory with the usual fields (higher-spin fields and matter). Already in the gravity case one can drop one $sl_2$ and get just a holomorphic (chiral) copy of the Virasoro algebra. For higher spins the smallest algebra seems to be $gl_\lambda \otimes \mathrm{Mat}_2$.\footnote{Exactly the algebra of this type has already appeared in \cite{Gaberdiel:2013vva}.} One can interpret it as the chiral half and get a color extension of the $W_\lambda$-algebra as asymptotic symmetries. In other words, the matter-coupled higher-spin gravity can be dual to a chiral CFT.

Auxiliary fields can also be introduced in higher dimensions, but do not get entangled with the physical fields and can be safely set to zero \cite{Vasiliev:1999ba,Kessel:2015kna,Sharapov:2019vyd}. Nevertheless, an interesting question is what would be the CFT interpretation of a higher-spin gravity where physical fields interact with auxiliary ones. Some examples with auxiliary fields of this type were given in \cite{Sharapov:2019vyd}. Thinking of the fundamental representation $V$ of a higher-spin algebra $\mathfrak{g}$ in $d>3$, $V\otimes V$ gives the physical degrees of freedom and $V\otimes V^*\sim \mathrm{End}(V)$ is $\mathfrak{g}$. The field-theoretical interpretation of taking $V^*$ on the CFT side is the inversion map. Therefore, a theory with both the higher-spin fields (dual to single trace operators described by $V\otimes V$) and auxiliary fields (taking values in $V\otimes V^*$) can have an interpretation of adding some non-local operators to local ones. 

Another possibility of getting a matter-coupled higher-spin gravity is to try to reconstruct it from the correlation functions of $W_\lambda$ minimal models. It is unclear whether the arguments regarding the nonlocality of the bulk vertices \cite{Bekaert:2015tva,Maldacena:2015iua,Sleight:2017pcz,Ponomarev:2017qab} apply to $AdS_3/CFT_2$ and how they depend on $\lambda$. It would also be important to check that. One possibility is that the bulk theory has to be very non-local (as in higher dimensions) but this is not captured by the higher-spin algebra cohomology which also mods out by nonlocal field redefinitions that have the form of formal series in derivatives of the matter fields. Two other loopholes are: (a) introducing the Maxwell field under which the matter is not charged, which requires $\lambda=1$; (b) flat space without an abelian spin-one and with/without the Maxwell field.\footnote{This way one misses all $AdS/CFT$ applications, of course.}  

On a different note, in $(A)dS$ but not in the flat space, one can argue that since all stress-tensors are removable by non-local (formal) field-redefinitions \cite{Prokushkin:1999xq}, the deformation need to be sought for in the class of `local' cocycles\footnote{Term `local' just means some class that is more narrow than the usual formal series in $p_{ij}=\epsilon^{AB} \pl_A^i \pl_B^j$. } modulo `local' redefinitions. Here, we can stress the difference between $3d$ and $d>3$. In $d>4$, already the free Fronsdal equations require $\mathcal{V}(h,h,C)$, ($h$ is the vielbein) which needs to be understood as the `free' limit of $\mathcal{V}(\omega,\omega,C)$. Therefore, the onset of the higher structure maps can be seen already at the free level. On the contrary, in $3d$ one does not expect any $\mathcal{V}(\omega,\omega,C)$ because there is no nontrivial $\mathcal{V}(h,h,C)$ (higher-spin fields are topological) and the stress-tensors $\mathcal{V}(h,h,C,C)$ belong to $\mathcal{V}(\omega,\omega,C,C)$. Now, $\mathcal{V}(\omega,\omega,C)$ is local (finite number of derivatives per spin) and $\mathcal{V}(\omega,\omega,C,C)$ can be nonlocal (since $C$ contains all derivatives of the matter fields and two $C$s can form an infinite series in derivatives). Therefore, the triviality of $\mathcal{V}(\omega,\omega,C)$ does not depend on the locality assumptions, but the absence of $\mathcal{V}(\omega,\omega,C,C)$ might.

At the technical level, it appears obvious that the $A_\infty$-algebra $\mathbb{A}$ of Chiral theory can be deformed in such a way that the underlying vector space is built on $A=gl_\lambda$. The $A_\infty$-algebra is of the pre-Calabi--Yau type, and hence, its underlying vector space is $A\oplus A^*$. It would be interesting to give an explicit description of the vertices for $\lambda\neq1/2$. It is also interesting to see what happens for the critical points of $\lambda=N\in \mathbb{Z}$, where it develops a two-sided ideal with the quotient being $sl(N)$. Since we cannot couple matter to just $sl(N)$ Chern--Simons fields, the ideal cannot be quotiented through the vertices. Nevertheless, this would give some extension of the $sl(N)$ matter-coupled theory, most likely with topologically massive fields along the lines of \cite{Boulanger:2014vya}, again along the lines of 'extend the multiplet till it works'.

\section*{Acknowledgments}
\label{sec:Aknowledgements}
We are also grateful to Thomas Basile, Xavier Bekaert, Nicolas Boulanger, Andrea Campoleoni, Stefan Fredenhagen, Matthias Gaberdiel and Daniel Grumiller for useful discussions and comments. This project has received funding from the European Research Council (ERC) under the European Union’s Horizon 2020 research and innovation programme (grant agreement No 101002551).

\appendix

\section{Flat limit}
\label{app:limit}
In order to take the flat limit of the free equations on $AdS_3$ with the algebra given by the even subalgebra of the Weyl algebra $A_1$, i.e. $gl_{1/2}$, we define
\begin{subequations}
    \begin{align}
        &V\left(\omega_1,\omega_2\right)=\frac1\Lambda\exp\left(p_{01}+p_{02}\right)\sinh\left(\Lambda p_{12}\right)\,,\\
        &V\left(\omega_1,e_2\right)=\frac1\Lambda\exp\left(p_{01}+p_{02}\right)\sinh\left(\Lambda p_{12}\right)\,,\\
        &V\left(e_1,e_2\right)=\Lambda\exp\left(p_{01}+p_{02}\right)\sinh\left(\Lambda p_{12}\right)\,,\\
        &U\left(\omega_1, C\right)=\frac1\Lambda\exp\left(p_{12}+p_{02}\right)\sinh\left(\Lambda p_{01}\right)\,,\\
        &U\left(e_1,C\right)=\exp\left(p_{12}+p_{02}\right)\cosh\left(\Lambda p_{01}\right)\,.
    \end{align}
\end{subequations}
It is important to remember that the fields are even functions of $y$. Therefore, certain components of the maps here-above vanish.

\section{Formal deformations: adjoint vs. bi-fundamental}
\label{app:Hoch}
It might be instructive to compare the two situations: a zero-form $C$ in the adjoint of some (higher-spin) algebra $A$ and a zero-form $C$ in the bi-fundamental of the same $A$. As usual, we assume that $A$ is an associative algebra. The two systems corresponding to the two types of representations where $C$ takes values are
\begin{align}
dC&=A\star C-C\star A &
dA&=A\star A\,,
\end{align}
and
\begin{align}
dC&=A\star C-C\star B\,, &
dA&=A\star A\,, &
dB&=B\star B\,.
\end{align}
We look for the first-order deformation of the two systems\footnote{In the derivation one needs to either assume that $A$ can be replaced by $A\otimes \mathrm{Mat}_N$ for any $N$ or just consider the terms with different order of arguments as different as a simplifying assumption. Mathematically, the vertices form an $L_\infty$-algebra, but under certain assumptions one can prove that it has to come as the symmetrization of some $A_\infty$-algebra's maps. }
\besubeqs
\begin{align}
dA&=A\star A+V_1(A,A,C)+V_2(A,C,A)+V_3(A,A,C)+\ldots \,,\\
dC&=A\star C-C\star A + \ldots
\end{align}
\esubeqs
and
\besubeqs
\begin{align}
dA&=A\star A+V_1(A,A,C)+\ldots\,,\\
dB&=B\star B+\tilde V_1(A,A,C)+\ldots\,,\\
dC&=A\star C-C\star B+\ldots\,.
\end{align}
\esubeqs
Here, we have displayed only some of the terms  that  contribute to the lowest order. 

It is easy to construct the vertices for the first system provided the associative algebra $A$ admits a nontrivial  deformation \cite{Sharapov:2019vyd}.
To the lowest order we can take
\begin{align}\label{hochsol}
    V_1(A,A,C)&= \phi_1(A,A)\star C\,, && V_{2,3}=0\,,
\end{align}
where $\phi_1$ is a nontrivial Hochschild cocycle that represent an element of $HH^2(A,A)$. Indeed, the equation for $V_{1}$ reads
\begin{align}
    0&=V_1(A\star A,A,C)-A\star V_1(A,A,C)+V_1(A,A,A\star C)-V_1(A,A\star A,C)
\end{align}
and is easily seen to be solved by \eqref{hochsol}. The only other equation involving $V_1$ is 
\begin{align}
0&=V_1(A,A,C)\star A-V_1(A,A,C\star A)
\end{align}
(at $V_2=0$) and is satisfied too. On the contrary, a similar system for the bi-fundamental representation is
\besubeqs
\begin{align}
0&=V_1(A\star A,A,C)-A\star V_1(A,A,C)+V_1(A,A,A\star C)-V_1(A,A\star A,C)\,,\\
0&=V_1(A,A,C)\star A+V_2(A\star A,C,A)-A\star V_2(A,C,A)-V_2(A,A\star C,A)\,,\\
0&=V_2(A,C,A)\star A-V_2(A,C,A\star A)-A\star V_3(C,A,A)+V_3(A\star C,A,A)\,,\\
0&=V_3(C,A,A)\star A-V_3(C,A,A\star A)+V_3(C,A\star A,A)\,,
\end{align}
\esubeqs
where we see `less' terms as compared to the adjoint representation case because some $A$ need to replaced by $B$ and hence contribute to other equations (with other orderings of $A$, $B$ and $C$). While the first equation can be solved as before $V_1(A,A,C)= \phi_1(A,A)\star C$, we cannot set $V_{2,3}=0$ and $V_2(A,C,A)=\phi_1(A,C)\star A$, $V_3=0$. However, this deformation is trivial, i.e. can be removed by a field redefinition. 

The discussion above does not prove the absence of a nontrivial deformation, but gives plausible arguments that it is not there or, at least, the procedure applicable to all the other higher-spin algebras does not work here. Strictly speaking, one has to study the cohomology of a given higher-spin algebra with values in the module of interest, which is a hard problem in general. 

Lastly, we perform the numerical study of the $V(A,A,C)$-vertex with the fields $A$ and $C$ taking values in
\begin{itemize}
    \item the adjoint of $A_1$ and the twisted-adjoint representation of $A_1$,
    \item the adjoint of $A_1^e$ for both,
    \item the adjoint of $A_1^e/\mathbb{C}$ for $A$ and  $A_1^e$ for $C$ that is a module under the commutator action of $A_1^e/\mathbb{C}$.  
    \end{itemize}
The obtained results suggest that the (anti-symmetrized and projected onto the even part $A_1^e$ or $A_1^e/\mathbb{C}$ if needed) factorized solution \eqref{hochsol} is still a representative of a single nontrivial cohomology class. Therefore, in all these cases the Hochschild cohomology captures the complete Chevalley--Eilenberg cohomology.

\footnotesize
\providecommand{\href}[2]{#2}\begingroup\raggedright\endgroup

\end{document}